\documentclass[twocolumn]{aastex62}
\usepackage{amsmath}

\shorttitle{Gas metallicity relations in galaxies with and without RPS}
\shortauthors{Franchetto et al.}

\usepackage{soul,xcolor}
\usepackage{comment}

\begin{document}
\setstcolor{cyan}
\newcommand\boldc[1]{\textcolor{cyan}{\textbf{#1}}}
\title{GASP XXVII: Gas-phase metallicity scaling relations in disk galaxies with and without ram-pressure stripping}

\author[0000-0001-9575-331X]{Andrea Franchetto}
\affiliation{Dipartimento di Fisica e Astronomia ``Galileo Galilei'', Universit\`a di Padova, vicolo dell'Osservatorio 3, IT-35122, Padova, Italy}
\affiliation{INAF - Astronomical Observatory of Padova, vicolo dell'Osservatorio 5, IT-35122 Padova, Italy}

\author[0000-0003-0980-1499]{Benedetta Vulcani}
\affiliation{INAF - Astronomical Observatory of Padova, vicolo dell'Osservatorio 5, IT-35122 Padova, Italy}

\author[0000-0001-8751-8360]{Bianca M. Poggianti}
\affiliation{INAF - Astronomical Observatory of Padova, vicolo dell'Osservatorio 5, IT-35122 Padova, Italy}

\author[0000-0002-7296-9780]{Marco Gullieuszik}
\affiliation{INAF - Astronomical Observatory of Padova, vicolo dell'Osservatorio 5, IT-35122 Padova, Italy}

\author[0000-0003-2589-762X]{Matilde Mingozzi}
\affiliation{INAF - Astronomical Observatory of Padova, vicolo dell'Osservatorio 5, IT-35122 Padova, Italy}

\author[0000-0002-1688-482X]{Alessia Moretti}
\affiliation{INAF - Astronomical Observatory of Padova, vicolo dell'Osservatorio 5, IT-35122 Padova, Italy}

\author[0000-0002-8238-9210]{Neven Tomi\v{c}i\'c}
\affiliation{INAF - Astronomical Observatory of Padova, vicolo dell'Osservatorio 5, IT-35122 Padova, Italy}

\author[0000-0002-7042-1965]{Jacopo Fritz}
\affiliation{Instituto de Radioastronom\'ia y Astrof\'isica, UNAM, Campus Morelia, A.P.\ 3-72, C.P.\ 58089, Mexico}

\author[0000-0002-4158-6496]{Daniela Bettoni}
\affiliation{INAF - Astronomical Observatory of Padova, vicolo dell'Osservatorio 5, IT-35122 Padova, Italy}

\author[0000-0003-2150-1130]{Yara L. Jaff\'e}
\affiliation{Instituto de Fisica y Astronomia, Universidad de Valparaiso, Avda. Gran Bretana 1111 Valparaiso, Chile}

\correspondingauthor{Andrea Franchetto}
\email{andrea.franchetto@phd.unipd.it}

\begin{abstract}

Exploiting the data from the GAs Stripping Phenomena in galaxies with MUSE (GASP) survey, we study the gas-phase metallicity scaling relations of a sample of 29 cluster galaxies undergoing ram-pressure stripping and of a
reference sample of (16 cluster and 16 field) galaxies with no significant signs of gas disturbance. We adopt the {\sc pyqz} code to infer the mean gas metallicity at the effective radius and achieve a well-defined mass-metallicity relation (MZR) in the stellar mass range $10^{9.25}\le M_\star \le 10^{11.5}\,{\rm M_\odot}$ with a scatter of 0.12~dex.
At any given mass, reference cluster and stripping galaxies have similar metallicities, while the field galaxies with  $M_\star < 10^{10.25}\,{\rm M_\odot}$ show on average lower gas metallicity than  galaxies in clusters.
Our results indicate that at the effective radius the chemical properties of the stripping galaxies are  independent of the ram-pressure stripping mechanism. Nonetheless, at the lowest masses we detect 4 stripping galaxies well above the common MZR that suggest a more complex scenario.
Overall, we  find signs of an anti-correlation between the metallicity and both the star formation rate and the galaxy size, in agreement with previous studies. No significant trends are instead found with the halo mass, clustercentric distance and local galaxy density in clusters.
In conclusion, we advise
a more detailed analysis of the spatially resolved gas metallicity maps of the galaxies, able to highlight effects of gas redistribution inside the disk due to the ram-pressure stripping.

\end{abstract}

\keywords{galaxies}

\section{Introduction}\label{sec:intro}

The gas-phase metallicity is known to show a well-established relation with the galaxy stellar mass in the range from $10^{6}$ to $10^{12}$~M$_\odot$ \citep{tremonti2004,kewley2008,mannucci2010,sanchez2013,barrera2017,hirschauer2018,sanchez2019,blanc2019}.
Decades of studies proposed several interpretations for this correlation invoking outflows of enriched gas driven by stellar/Active Galactic Nuclei (AGN) feedback \citep[e.g.][]{garnett2002,derossi2017,chisholm2018}, infall of pristine gas \citep{mannucci2010}, evolutionary stage and downsizing (high-mass galaxies evolve more rapidly than low-mass galaxies, becoming metal-rich at earlier epochs; \citealt{maiolino2008}), and dependence of the initial mass function (IMF) on galaxy mass with changes of the stellar yields \citep{demasi2018}. However, the shape of the mass-metallicity relation (MZR) could stem from the combination of all these processes \citep{maiolino2019}. Recent studies at low and intermediate redshift investigated the role of several parameters to explain the scatter around the relation and discern between the mechanisms that are shaping the relation.

In the last decade the attention turned on the role of star formation rate (SFR) in explaining the scatter along the MZR. Using a sample of $\sim$40\,000 SDSS star-forming galaxies, \citet{ellison2008} was the first to find an anti-correlation between the gas-phase metallicity and the specific SFR (sSFR, SFR per unit of galaxy stellar mass) at a given mass.
The existence of a well-defined relation between the stellar mass, gas-phase metallicity and SFR was discussed in detail by \citet{mannucci2010} exploiting a sample of $\sim$140\,000 star-forming galaxies from SDSS. The analytic form of their so-called Fundamental Metallicity Relation (FMR) describes a surface in the 3D space of the involved parameters and the scatter in metallicity along this surface is reduced with respect to that observed for the MZR
\citep[see also][]{laralopez2010,hunt2012,yates2012}.
The explanation of the FMR invokes the metal-poor gas accretion that, on one side, implicates metal dilution and, on the other side, produces star-formation activity \citep{mannucci2010}.
Integral-field unit data seem to be still consistent with the mass-SFR-metallicity relation \citep{cresci2019}, even if the results are largely debated and some authors find a weak or no secondary dependence on SFR \citep{sanchez2013,sanchez2017,sanchez2019,barrera2017}.

\citet{ellison2008} also explored the connection between the stellar mass, the gas-phase metallicity and the galaxy size and found that compact galaxies exhibit higher metallicity than larger ones. \citet{sanchezalmeida2018} investigated this relation using the EAGLE cosmological numerical simulation and discovered similar results. Their analysis shows that the anti-correlation between the gas-phase metallicity and the galaxy size is due to the late gas accretion. Galaxies grow in size with time, so if they experience recent metal-poor infall of gas they will be bigger and with lower gas-phase metallicity than those formed earlier.

Many studies show that also the environment may play a role in shaping the MZR and that the over-dense environment, where many specific  mechanisms can affect the gas reservoir, can alter the gas-phase metallicity. 
For example, \citet{ellison2009} studied a sample of $\sim$1300 cluster galaxies and reported that on average they have an overabundance of gas-phase metallicity up to 0.05~dex in comparison with a control sample of galaxies that are not cluster members. However, their study ascribes this effect to the local galaxy density and not to the cluster membership, as control galaxies at locally high densities exhibit similar metal-enhancements to the cluster ones. 
In addition, for the massive cluster galaxies the overabundance resulted independent of the global cluster properties (e.g. virial radius, halo mass). 
\citet{peng2014} analyzed $\sim$16\,000 satellite galaxies and observed a strong correlation between the gas metallicity and the over-density at a given mass, proposing metal-enriched inflows in crowded environments as explanation.
\citet{pilyugin2017}, using a sample of $\sim$77\,600 late-type galaxies, found that mainly low mass galaxies ($10^{9.1}<M_\star<10^{9.6}$~M$_\odot$) show on average an excess of gas metallicity in the densest environments, but a large scatter is observed at any density of the environment.
The dependence on the environment could be explained in terms of gas content. In fact, the metallicity  anti-correlates with the SFR and the gas fraction, that in turn appear to be both anti-correlated with the local density \citep{wu2017}. \citet{maier2019a,maier2019b} found an enhancement of oxygen abundance for galaxies inside the virialized region of clusters, and argued for the importance of environmental processes such as strangulation (halo gas removal by the intracluster medium -ICM- interaction, \citealt{larson1980}).

All the aforementioned studies have therefore highlighted the connection between SFR, galaxy size and environment with the gas content of the galaxies and simulations indeed find a strong relation between the gas-phase metallicity and the gas fraction of galaxies \citep{derossi2016,derossi2017}.
In the context of  galaxy evolution, testing the metallicity scaling relations provides an important tool to study the many physical processes in galaxies and, in particular, the effects due to the mechanisms affecting the gas reservoir.
A gas deficit is often observed in cluster galaxies due to different environmental processes \citep{boselli2006}: thermal evaporation \citep{cowie1977}, starvation \citep{larson1980}, ram-pressure stripping (RPS; \citealt{gunn1972}).
RPS due to the interaction between the ICM and the interstellar medium (ISM) is one of the most efficient gas removal processes from galaxies in clusters. Indeed, the study of the gas-phase metallicity in galaxies undergoing this process could provide constraints on the gas redistribution inside the galaxy disk.

In this paper we indeed explore the gas metallicity scaling relations of a sample of galaxies undergoing RPS, exploiting the Gas Stripping Phenomena in galaxies with MUSE (GASP; \citealt{poggianti2017}) data. GASP is an ESO Large Program carried out with the integral-field spectrograph MUSE \citep{bacon2010} mounted on the VLT (Paranal) aiming at systematically studying
the gas removal processes from galaxies in different environments. MUSE allow to explore in detail the spatially resolved distribution of the ionized gas emission not only within the galaxy disk but also along the gas stripped beyond the stellar extent.

So far, the gas-phase metallicity was derived only for a limited number of extreme RPS galaxies \citep[e.g.][]{fossati2016,poggianti2017,gullieuszik2017,bellhouse2017,bellhouse2019,moretti2018a}, in this work we instead analyze this quantity for the first time in a statistically meaningful sample,
and compare the  properties of RPS galaxies to those of a sample of undisturbed galaxies in clusters and field from the same survey.

In Sect.~\ref{sec:sample} we present the galaxy sample extracted from the GASP data. Section~\ref{sec:method} describes the methods adopted to measure the main properties of the galaxies. The MZRs of the sample are shown and analyzed in Sect.~\ref{sec:result}, while the interconnection between the gas-phase metallicity and other parameters is explored in detail in Sect.~\ref{sec:discus}. In Sect.~\ref{sec:concl} we conclude with a summary of our work.

This paper is the first of a series focusing on the statistical study of the chemical properties of the ionized gas component in galaxies experiencing RPS.

This paper adopts a \citet{chabrier2003} initial mass function (IMF) and standard concordance cosmology parameters $H_0=70~{\rm km~s^{-1}~Mpc^{-1}}$ , $\Omega_M=0.3$ and $\Omega_\Lambda=0.7$.

\section{Sample and observations}\label{sec:sample}
The GASP project observed 114 disk galaxies at $0.04<z<0.07$ in different environments (galaxy clusters, groups, filaments and isolated) and with stellar mass in the range $10^{8.7}<M_\star<10^{11.5}~{\rm M_\odot}$.
The sample includes 76 galaxies in clusters taken from the WINGS \citep{fasano2006} and OMEGAWINGS \citep{gullieuszik2015} cluster surveys and 38 galaxies in less massive environments taken from the PM2GC catalogue \citep{calvi2011}. 
The GASP targets and the observing strategy are described in detail in \citet{poggianti2017}. 

The final datacubes consist in $300\times 300$ spaxels with a spatial sampling of 
$0.2''\times0.2''$.
We stress that for all GASP galaxies the FoV of MUSE is able to cover from 3 to 15 effective radii ($R_{\rm e}$) from their center, with a mean of $7\,R_{\rm e}$.
This coverage allows us to observe the full optical extent of the galaxies and includes a wide portion of sky around them.

\subsection{Selection of the sample}

We extract from the GASP sample the cluster galaxies showing unilateral ionized gas beyond their stellar disk ( from few kiloparces to 100~kpc), while having the old stellar component (formed more than $6\times 10^8$ years ago) morphologically undisturbed. These signs indicate galaxies are suffering from ram-pressure by the ICM.
We include galaxies at different stripping stages (from initial stripping with a lopsided distribution of the gas component to extreme stripping galaxies with tens-kiloparsec gas tails) and exclude truncated disk galaxies (galaxies with gas disk less extended than the stellar one and with no ionized gas tails). This selection yields 29 galaxies that we will call ``stripping'' sample.

From the GASP sample, we also draw a sample of galaxies which will constitute a ``reference'' sample. These galaxies, located both in clusters and in the less dense environments (field and groups), do not exhibit clear signs of ongoing gas stripping processes (no evident gas tails or gas debris well-beyond the stellar disk) and have regular ionized gas and stellar distributions.
Nonetheless, we note that these galaxies might still be partially affected by some physical processes that we are not able to identify.\footnote{We have however removed from the sample peculiar galaxies showing signs of tidal interactions,  galaxies with a companions,
and the field galaxies with clear signs of specific processes on act
\citet{vulcani2018a,vulcani2018c,vulcani2019a}.}

The reference sample includes 16 cluster galaxies and 16 galaxies in less dense environments, defined as ``reference cluster''  and ``reference field'' sample, respectively.

The complete list of galaxies included in the analysis will be  given in Tab.~\ref{tab:controlsample}.

\begin{figure}
\centering
\includegraphics[width=\columnwidth]{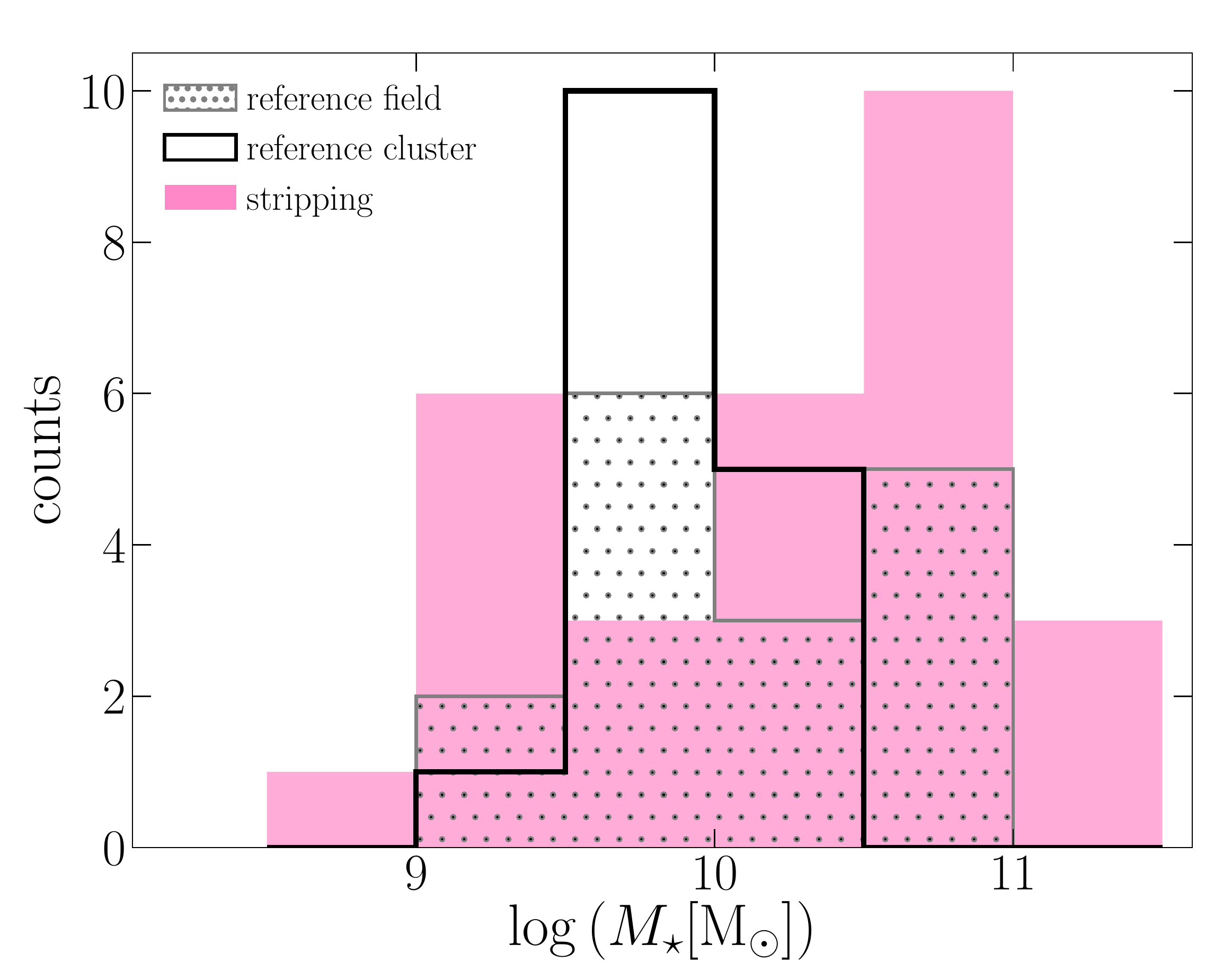}
\caption{Galaxy stellar mass distribution of the reference field (dotted gray histogram), reference cluster (solid black histogram) and stripping (pink histogram) samples.\label{fig:histm}}
\end{figure}

Figure~\ref{fig:histm} shows the stellar mass distributions of the different samples. The stripping galaxies span a wide range in stellar mass, going from $10^{8.7}$ to $10^{11.5}~{\rm M_\odot}$. Reference cluster galaxy masses range from $10^{9.2}$ to $10^{10.5}~{\rm M_\odot}$, while reference field galaxy masses reach $10^{11}~{\rm M_\odot}$. The mass ranges of the total reference sample (cluster+field) and the stripping sample overlap for about two orders of magnitude, even though, according to the Kolmogorov-Smirnov (KS) test, their mass distributions are drawn from different parent distributions
(p-value$<$5\%).

\begin{figure}
\includegraphics[width=\columnwidth]{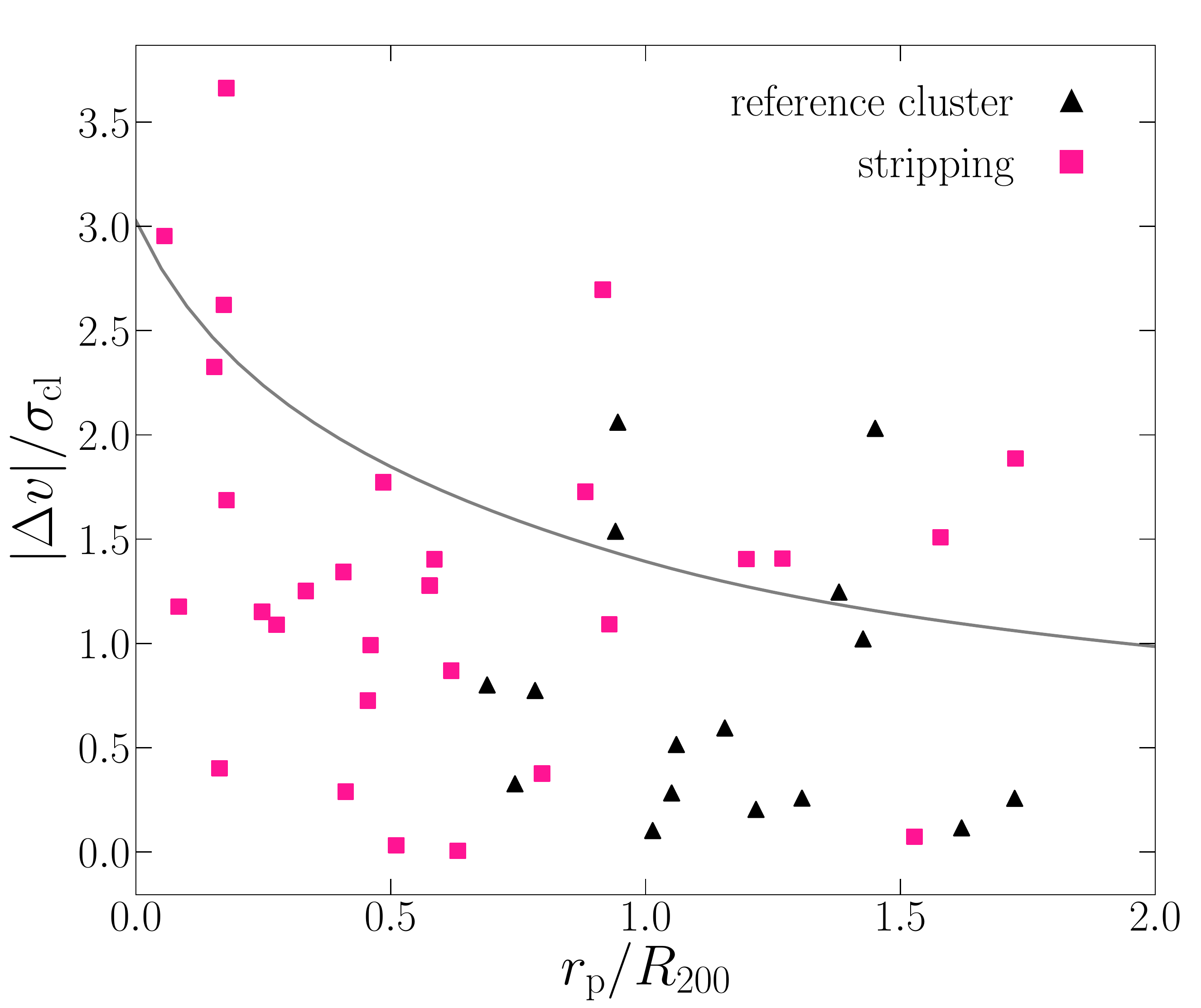}
\caption{Projected velocity vs.\ projected position phase-space diagram of the reference cluster galaxies (black triangles) and the stripping galaxies (pink squares). The curve shows the escape velocity in a \citet{navarro1996} halo assuming a concentration value of 6.\label{fig:pps}}
\end{figure}

Regarding the mass of the host environment, galaxies in the reference field sample belong to low-mass dark matter halo ($11.2<\log M_{\rm halo}/M_\sun<12.9$) with at most 5 members. Instead, the masses of the clusters, in which the stripping and the reference cluster galaxies live, span a range of $13.8<\log M_{\rm halo}/M_\sun<15.4$.\footnote{Halo masses for all the PM2GC sample are derived in \cite{paccagnella2019}, while those of the clusters are taken from \citet{biviano2017} and Munari et al.\ (in prep.).}

Considering only cluster galaxies, Fig.~\ref{fig:pps} shows the projected phase-space diagram for stripping and reference galaxies.
This diagram compares the line-of-sight velocity of each galaxy with respect to the cluster mean velocity $|\Delta v|$, normalized by the cluster velocity dispersion $\sigma_{\rm cl}$ against its projected clustercentric distance $r_{\rm p}$
normalized by the cluster virial radius $R_{200}$ ($\sigma_{\rm cl}$ and $R_{200}$ are taken from Gullieuszik et al.\ -submitted-). The phase-space diagram allows us to study galaxy properties as a function of the orbital histories of the galaxies within the clusters.
As shown in \citet{jaffe2018}, the stripping efficiency increases as galaxies move towards higher velocities and lower distances from the cluster cores. 
Galaxies of the two samples occupy different regions of the diagram. 24 stripping galaxies (83\% of the stripping sample) are within $1~R_{200}$, while the reference cluster galaxies are all located beyond 0.7 $R_{200}$. Although both samples show a large velocity scatter, the stripping sample has a higher mean value ($\langle|\Delta v|/\sigma_{\rm cl}\rangle=1.35$) than the reference cluster one ($\langle|\Delta v|/\sigma_{\rm cl}\rangle=0.76$). The two-dimensional KS test states with high confidence that the two samples are not drawn from the same parent distribution ($\text{p-value}<0.0004$). In Sec.~\ref{sec:discus} we explore the impact of this difference on the gas-phase metallicity of the sample galaxies.

\section{Data analysis}\label{sec:method}
The methods employed in the analysis of the MUSE datacubes are explained in detail in \citet{poggianti2017}.
Briefly,
data were reduced with the most updated version of the MUSE pipeline at the moment of observations \citep{bacon2010}\footnote{\url{http://www.eso.org/sci/software/pipelines/muse}} and corrected for extinction due to our own Galaxy assuming the extinction law from \citet{cardelli1989}. The datacubes were average filtered in the spatial direction with a 5~$\times$~5 pixel kernel, corresponding to $1"=0.7-1.3$~kpc, depending on the galaxy redshift.

Each spectrum is corrected for the underlying stellar absorption subtracting the stellar-only component derived with our spectrophotometric code {\sc sinopsis} \citep{fritz2017}.
The emission-line fluxes with the associated errors are measured by the IDL software {\sc kubeviz} \citep{fossati2016} that employs Gaussian profile fitting. The emission lines of interest for this work are: H$\beta$, [{\sc O\,iii}]5007, [{\sc O\,i}]6300, H$\alpha$, [{\sc N\,ii}]6583, [{\sc S\,ii}]6716 and [{\sc S\,ii}]6731. We adopt the \citet{cardelli1989} extinction law to correct the emission-line fluxes for extinction by dust internal to the galaxy considering the observed Balmer decrement and assuming an intrinsic H$\alpha$/H$\beta$ ratio of 2.86.

In addition, {\sc sinopsis} delivers several spatially resolved properties of the stellar component. In this paper we make use of the stellar masses
and calculate the total stellar mass of each galaxy summing of the values of all spaxels belonging to the galaxy disk, as in \citet{vulcani2018b}. The definition of the galaxy boundary is developed by Gullieuszik et al.\ (submitted) and already exploited by \citet{vulcani2018b} and \citet{poggianti2019}.

We use the diagnostic BPT-diagram \citep{baldwin1981} [{\sc O\,iii}]5007/H$\beta$ vs [{\sc N\,ii}]6583/H$\alpha$ to divide the regions dominated by star formation (SF), Composite (SF+AGN), AGN and LI(N)ERs (low-ionization nuclear emission regions). We use the separation lines by \citet{kauffmann2003}, \citet{kewley2001}, and \citet{sharp2010}.  We note that some of the most massive galaxies in the stripping sample host either an AGN or LINERs, while no galaxy of the reference sample does. A detailed analysis of the AGN in these galaxies is presented in \citet{radovich2019}.

\subsection{Structural parameters}\label{sec:strucparam}
The structural parameters
of each galaxy are derived from the analysis of the I-band images obtained from the integrated MUSE datacubes using the Cousins I-band filter response curve.
In Appendix~\ref{app:re} we describe in detail the surface brightness analysis carried out to derive all the quantities.
Briefly, we use the {\sc ellipse} task in {\sc iraf} to derive the isophotal segmentation of the galaxies \citep{jedrze1987}. The extraneous sources and bright spots are masked out to achieve a smooth luminosity distribution of the galaxy. The {\sc ellipse} algorithm fits the galaxy isophotes with a series of ellipses such to minimize the deviations from the true shape of the isophotes. It then returns the values of semi-major axes, surface brightness, position angle ($P\!A$) and ellipticity ($\varepsilon$) of each elliptical isophote. We calculate the luminosity growth curve $L(R)$ of the galaxies by the trapezoidal integration of their surface brightness profiles. By definition, the effective radius ($R_{\rm e}$) is the radius such as $L(R_{\rm e})/L_{\rm tot}=0.5$. We set $L_{\rm tot}$ as the total luminosity within the most external fitted isophote assuring us that it encloses entirely the full optical extent of the galaxy (see Appendix~\ref{app:re}).
We derived the $\varepsilon$ and
$P\!A$ of the disk as the average values of the elliptical isophotes corresponding to the galaxy disk.  The galaxy inclination $i$ is computed as
\begin{equation}
\cos^2i=\frac{(1-\varepsilon)^2-q_0^2}{1-q_0^2}
\end{equation}
where we assume an intrinsic flattening for galaxies of $q_0=0.13$ \citep{giovanelli1994}.

Note that two stripping galaxies (JO149 and JO95) have an irregular I-band morphology that prevents a good estimate of their structural parameters. Therefore,
we fix their $\varepsilon$ to zero and estimate their $R_{\rm e}$ using the mass-size relation of Eq.~\ref{eq:mrestrip}
in Appendix~\ref{app:re}
for the stripping galaxies.

With the quantities just described we deproject the position of each spaxel to derive its physical galactocentric distance. We do not apply the deprojection procedure for galaxies with low inclination ($i<35^\circ$), because the correction effect could be smaller than the uncertainties that the deprojection could introduce.

The structural parameters for all our galaxies are listed in Tab.~\ref{tab:controlsample}.

\begin{deluxetable*}{rrrrrrrrrrrr}
\tabletypesize{\footnotesize}
\tablecaption{Properties of the sample. Columns are: 1) GASP ID number; 2) sample (RF is reference field, RC is reference cluster, S is stripping); 3) galaxy redshift (for field galaxies) or mean galaxy cluster redshift (for cluster galaxies); 4) logarithm of the galaxy stellar mass; 5) $R_{\rm e}$; 6) disk inclination; 7) $P\!A$ measured counter-clockwise from North direction; 8), 9), 10) and 11) mean gas-phase metallicity within $0.1\,R_{\rm e}$, within $0.5\,R_{\rm e}$, at $R_{\rm e}$ and beyond $0.5\,R_{\rm e}$, respectively, using {\sc pyqz}; 12) mean gas-phase metallicity at $R_{\rm e}$ using the O3N2 calibration of \citet{curti2017}. \label{tab:controlsample}}
\tablehead{\colhead{ID} & \colhead{sample} &\colhead{redshift} & \colhead{$\log\,(M_\star/M_\odot)$} & \colhead{$R_{\rm e}$} & \colhead{$i$} & \colhead{$P\!A$} & \colhead{$\langle{\rm \frac{O}{H}}\rangle_{<0.1\,R_{\rm e}}$} & \colhead{$\langle{\rm \frac{O}{H}}\rangle_{<0.5\,R_{\rm e}}$} & \colhead{$\langle{\rm \frac{O}{H}}\rangle_{@\,R_{\rm e}}$} & \colhead{$\langle{\rm \frac{O}{H}}\rangle_{\rm disk}$} & \colhead{$\langle{\rm \frac{O}{H}}\rangle_{@\,R_{\rm e}}$CU17} \\
 &  & & \colhead{(dex)} & \colhead{(arcsec)} & \colhead{(degree)} & \colhead{(degree)} & \colhead{(dex)} & \colhead{(dex)} & \colhead{(dex)} & \colhead{(dex)} & \colhead{(dex)}}
\decimals
\startdata
          JO73 &  RF & 0.0713 &    10.03$\pm$0.10 &     4.24$^{+0.25}_{-0.27}$ &     33$\pm$1 &     75$\pm$2 &     8.98$\pm$0.01 &     8.86$\pm$0.08 &     8.59$\pm$0.06 &     8.56$\pm$0.13 &     8.59$\pm$0.05 \\
        P13384 & RF &  0.0512 &     9.85$\pm$0.11 &     3.37$^{+0.29}_{-0.28}$ &     29$\pm$8 &    171$\pm$38 &     8.82$\pm$0.00 &     8.76$\pm$0.04 &     8.56$\pm$0.04 &     8.53$\pm$0.09 &     8.56$\pm$0.05 \\
         P15703 & RF &   0.0423 &    11.00$\pm$0.08 &     5.08$^{+0.22}_{-0.23}$ &     66$\pm$1 &    154$\pm$1 &      \dots &      \dots &     9.15$\pm$0.02 &     9.09$\pm$0.08 &     8.78$\pm$0.02 \\
         P17945 & RF &  0.0439 &     9.75$\pm$0.10 &     4.08$^{+0.26}_{-0.28}$ &     42$\pm$6 &     40$\pm$7 &     8.79$\pm$0.01 &     8.72$\pm$0.04 &     8.54$\pm$0.04 &     8.52$\pm$0.10 &     8.55$\pm$0.04 \\
         P20769 & RF &   0.0489 &     9.45$\pm$0.10 &     2.55$^{+0.14}_{-0.15}$ &     57$\pm$4 &     54$\pm$5 &     8.43$\pm$0.01 &     8.44$\pm$0.03 &     8.34$\pm$0.06 &     8.28$\pm$0.15 &     8.52$\pm$0.03 \\
         P20883 & RF &   0.0614 &     9.91$\pm$0.09 &     4.24$^{+0.25}_{-0.27}$ &     57$\pm$3 &    166$\pm$3 &     8.92$\pm$0.01 &     8.87$\pm$0.04 &     8.67$\pm$0.04 &     8.60$\pm$0.12 &     8.63$\pm$0.02 \\
         P21734 & RF &   0.0685 &    10.80$\pm$0.09 &     6.76$^{+0.81}_{-0.78}$ &     32$\pm$4 &    111$\pm$28 &     9.20$\pm$0.01 &     9.19$\pm$0.03 &     8.99$\pm$0.07 &     8.81$\pm$0.25 &     8.72$\pm$0.04 \\
         P25500 & RF &   0.0604 &    10.82$\pm$0.09 &     8.12$^{+0.69}_{-0.76}$ &     50$\pm$7 &    105$\pm$9 &     9.13$\pm$0.01 &     9.12$\pm$0.03 &     9.03$\pm$0.04 &     8.86$\pm$0.18 &     8.76$\pm$0.03 \\
         P42932 & RF &   0.0410 &    10.51$\pm$0.07 &     5.75$^{+0.24}_{-0.24}$ &     40$\pm$3 &    130$\pm$4 &     9.19$\pm$0.01 &     9.16$\pm$0.02 &     9.11$\pm$0.02 &     9.04$\pm$0.08 &     8.79$\pm$0.01 \\
         P45479 & RF &   0.0515 &    10.57$\pm$0.07 &     5.24$^{+0.20}_{-0.21}$ &     53$\pm$2 &     32$\pm$1 &     9.15$\pm$0.01 &     9.15$\pm$0.02 &     9.07$\pm$0.03 &     8.99$\pm$0.10 &     8.77$\pm$0.02 \\
         P48157 & RF &   0.0615 &    10.59$\pm$0.08 &     5.95$^{+0.41}_{-0.43}$ &     49$\pm$3 &    115$\pm$9 &      \dots &     9.08$\pm$0.03 &     8.98$\pm$0.05 &     8.83$\pm$0.15 &     8.74$\pm$0.02 \\
         P57486 & RF &   0.0529 &     9.94$\pm$0.10 &     4.18$^{+0.26}_{-0.30}$ &     57$\pm$3 &    101$\pm$2 &     8.95$\pm$0.01 &     8.89$\pm$0.04 &     8.68$\pm$0.04 &     8.61$\pm$0.12 &     8.63$\pm$0.02 \\
         P63947 & RF &   0.0562 &     9.34$\pm$0.10 &     2.90$^{+0.19}_{-0.21}$ &     50$\pm$10 &    172$\pm$11 &     8.55$\pm$0.02 &     8.51$\pm$0.06 &     8.38$\pm$0.09 &     8.36$\pm$0.12 &     8.55$\pm$0.06 \\
           P648 & RF &   0.0660 &    10.43$\pm$0.09 &     5.37$^{+0.39}_{-0.40}$ &     45$\pm$4 &     47$\pm$8 &     9.03$\pm$0.01 &     8.97$\pm$0.04 &     8.80$\pm$0.05 &     8.74$\pm$0.14 &     8.68$\pm$0.03 \\
           P669 & RF &   0.0458 &    10.49$\pm$0.08 &     6.36$^{+0.32}_{-0.36}$ &     33$\pm$4 &    148$\pm$7 &      \dots &     9.14$\pm$0.04 &     9.04$\pm$0.06 &     9.02$\pm$0.08 &     8.77$\pm$0.03 \\
           P954 & RF &   0.0450 &     9.63$\pm$0.11 &     4.01$^{+0.29}_{-0.34}$ &     42$\pm$4 &    130$\pm$9 &     8.69$\pm$0.01 &     8.63$\pm$0.06 &     8.49$\pm$0.05 &     8.42$\pm$0.13 &     8.55$\pm$0.04 \\
A3128\_B\_0148 & RC &   0.0603 &     9.85$\pm$0.10 &     2.30$^{+0.10}_{-0.10}$ &     64$\pm$3 &    119$\pm$2 &     8.73$\pm$0.00 &     8.73$\pm$0.01 &     8.68$\pm$0.02 &     8.57$\pm$0.09 &     8.60$\pm$0.02 \\
A3266\_B\_0257 & RC &   0.0592 &     9.92$\pm$0.10 &     2.87$^{+0.09}_{-0.10}$ &     66$\pm$1 &     23$\pm$1 &     8.94$\pm$0.00 &     8.91$\pm$0.03 &     8.79$\pm$0.04 &     8.68$\pm$0.11 &     8.68$\pm$0.01 \\
 A3376\_B\_0261 & RC &   0.0465 &    10.53$\pm$0.08 &     4.51$^{+0.29}_{-0.29}$ &     19$\pm$5 &    141$\pm$48 &     9.08$\pm$0.01 &     9.12$\pm$0.02 &     9.05$\pm$0.03 &     8.89$\pm$0.15 &     8.77$\pm$0.01 \\
  A970\_B\_0338 & RC &   0.0587 &    10.06$\pm$0.08 &     3.86$^{+0.21}_{-0.17}$ &     75$\pm$1 &      4$\pm$2 &     8.99$\pm$0.01 &     8.92$\pm$0.05 &     8.81$\pm$0.03 &     8.73$\pm$0.09 &     8.69$\pm$0.02 \\
          JO102 & RC &   0.0603 &    10.01$\pm$0.08 &     2.86$^{+0.21}_{-0.22}$ &     71$\pm$2 &     33$\pm$1 &     9.01$\pm$0.00 &     9.00$\pm$0.01 &     8.95$\pm$0.04 &     8.89$\pm$0.07 &     8.73$\pm$0.02 \\
          JO128 & RC &   0.0548 &     9.90$\pm$0.11 &     6.69$^{+0.57}_{-0.66}$ &     23$\pm$7 &     95$\pm$24 &     8.85$\pm$0.03 &     8.73$\pm$0.08 &     8.51$\pm$0.09 &     8.50$\pm$0.13 &     8.56$\pm$0.06 \\
          JO138 & RC &   0.0554 &     9.65$\pm$0.15 &     4.15$^{+0.32}_{-0.34}$ &     75$\pm$3 &      1$\pm$3 &     8.45$\pm$0.01 &     8.42$\pm$0.05 &     8.30$\pm$0.07 &     8.28$\pm$0.12 &     8.54$\pm$0.04 \\
          JO159 & RC &   0.0483 &     9.82$\pm$0.11 &     3.81$^{+0.36}_{-0.39}$ &     40$\pm$6 &     70$\pm$7 &     8.56$\pm$0.01 &     8.47$\pm$0.05 &     8.37$\pm$0.06 &     8.37$\pm$0.10 &     8.48$\pm$0.06 \\
           JO17 & RC &   0.0447 &    10.16$\pm$0.08 &     6.78$^{+0.49}_{-0.52}$ &     60$\pm$2 &     64$\pm$3 &     9.07$\pm$0.01 &     9.01$\pm$0.04 &     8.86$\pm$0.04 &     8.83$\pm$0.09 &     8.73$\pm$0.03 \\
          JO180 & RC &   0.0623 &     9.98$\pm$0.10 &     3.56$^{+0.20}_{-0.21}$ &     26$\pm$9 &    151$\pm$31 &     9.09$\pm$0.00 &     9.04$\pm$0.03 &     8.90$\pm$0.04 &     8.90$\pm$0.07 &     8.74$\pm$0.02 \\
          JO197 & RC &   0.0545 &    10.04$\pm$0.10 &     4.14$^{+0.20}_{-0.22}$ &     57$\pm$4 &      3$\pm$4 &     9.00$\pm$0.00 &     8.95$\pm$0.04 &     8.75$\pm$0.03 &     8.69$\pm$0.11 &     8.67$\pm$0.03 \\
          JO205 & RC &   0.0489 &     9.52$\pm$0.11 &     3.66$^{+0.21}_{-0.25}$ &     55$\pm$3 &    139$\pm$6 &     8.64$\pm$0.01 &     8.59$\pm$0.04 &     8.46$\pm$0.03 &     8.40$\pm$0.11 &     8.54$\pm$0.02 \\
           JO41 & RC &   0.0464 &    10.20$\pm$0.07 &     5.08$^{+0.48}_{-0.48}$ &     26$\pm$6 &     77$\pm$22 &     9.10$\pm$0.01 &     9.11$\pm$0.04 &     9.04$\pm$0.06 &     9.04$\pm$0.08 &     8.79$\pm$0.03 \\
           JO5 & RC &   0.0653 &    10.27$\pm$0.10 &     3.88$^{+0.25}_{-0.26}$ &     44$\pm$4 &     24$\pm$3 &     9.13$\pm$0.01 &     9.10$\pm$0.02 &     8.98$\pm$0.03 &     8.75$\pm$0.22 &     8.73$\pm$0.02 \\
           JO68 & RC &  0.0579 &     9.99$\pm$0.09 &     4.40$^{+0.25}_{-0.28}$ &     56$\pm$1 &     66$\pm$2 &     9.04$\pm$0.02 &     8.97$\pm$0.04 &     8.83$\pm$0.05 &     8.79$\pm$0.09 &     8.69$\pm$0.02 \\
           JO89 &  RC & 0.0419 &     9.73$\pm$0.08 &     6.41$^{+0.57}_{-0.61}$ &     66$\pm$1 &     80$\pm$2 &     8.83$\pm$0.02 &     8.79$\pm$0.06 &     8.70$\pm$0.11 &     8.69$\pm$0.12 &     8.65$\pm$0.04 \\
        JO113 & S &  0.0595 &     9.69$\pm$0.09 &     3.53$^{+0.19}_{-0.23}$ &     73$\pm$1 &     22$\pm$1 &     8.76$\pm$0.00 &     8.74$\pm$0.02 &     8.61$\pm$0.06 &     8.52$\pm$0.10 &     8.58$\pm$0.05 \\
        JO135 & S &  0.0554 &    10.99$\pm$0.07 &     4.61$^{+0.15}_{-0.17}$ &     64$\pm$2 &     43$\pm$1 &      \dots &      \dots &      \dots &     8.98$\pm$0.09 &      \dots \\
        JO141 & S &   0.0554 &    10.68$\pm$0.13 &     5.39$^{+0.32}_{-0.36}$ &     70$\pm$2 &     71$\pm$3 &      \dots &     9.11$\pm$0.02 &     9.02$\pm$0.05 &     8.88$\pm$0.14 &     8.73$\pm$0.02 \\
        JO144 & S &   0.0480 &    10.51$\pm$0.14 &     3.83$^{+0.13}_{-0.15}$ &     66$\pm$3 &     29$\pm$4 &     9.05$\pm$0.01 &     9.04$\pm$0.02 &     8.96$\pm$0.03 &     8.90$\pm$0.08 &     8.71$\pm$0.02 \\
        JO147 & S &   0.0483 &    11.03$\pm$0.08 &     8.42$^{+0.59}_{-0.83}$ &     82$\pm$2 &     53$\pm$1 &      \dots &     9.18$\pm$0.01 &     9.15$\pm$0.03 &     9.11$\pm$0.04 &     8.79$\pm$0.03 \\
        JO149 & S &   0.0483 &     8.76$\pm$0.15 &      2.66 &      0 &      0 &     8.10$\pm$0.01 &     8.01$\pm$0.07 &     8.16$\pm$0.18 &     8.12$\pm$0.21 &     8.32$\pm$0.05 \\
        JO160 & S &   0.0483 &    10.06$\pm$0.11 &     5.36$^{+0.48}_{-0.54}$ &     59$\pm$2 &     87$\pm$7 &     9.14$\pm$0.01 &     9.06$\pm$0.05 &     8.79$\pm$0.06 &     8.74$\pm$0.12 &     8.65$\pm$0.04 \\
        JO162 & S &   0.0492 &     9.42$\pm$0.10 &     3.97$^{+0.20}_{-0.21}$ &     72$\pm$4 &    150$\pm$3 &     8.81$\pm$0.00 &     8.80$\pm$0.03 &     8.74$\pm$0.08 &     8.65$\pm$0.10 &     8.67$\pm$0.03 \\
        JO171 & S &   0.0553 &    10.61$\pm$0.06 &    12.07$^{+2.03}_{-2.67}$ &     18$\pm$0 &    113$\pm$42 &      \dots &     8.93$\pm$0.07 &     8.91$\pm$0.05 &     8.89$\pm$0.07 &     8.69$\pm$0.03 \\
        JO175 & S &   0.0460 &    10.50$\pm$0.07 &     3.61$^{+0.41}_{-0.38}$ &     44$\pm$4 &     56$\pm$3 &     9.24$\pm$0.00 &     9.22$\pm$0.02 &     9.08$\pm$0.02 &     8.91$\pm$0.15 &     8.77$\pm$0.01 \\
        JO181 & S &   0.0579 &     9.09$\pm$0.12 &     2.97$^{+0.33}_{-0.30}$ &     66$\pm$3 &    136$\pm$7 &     8.17$\pm$0.01 &     8.17$\pm$0.05 &     8.07$\pm$0.10 &     8.10$\pm$0.16 &     8.35$\pm$0.07 \\
        JO194 & S &   0.0488 &    11.18$\pm$0.09 &    10.82$^{+0.69}_{-1.01}$ &     39$\pm$4 &     60$\pm$14 &      \dots &     \dots &     9.17$\pm$0.04 &     9.15$\pm$0.04 &     8.80$\pm$0.02 \\
        JO200 & S &   0.0557 &    10.82$\pm$0.07 &     9.06$^{+0.64}_{-0.70}$ &     46$\pm$3 &     10$\pm$6 &      \dots &     9.20$\pm$0.03 &     9.13$\pm$0.04 &     8.97$\pm$0.17 &     8.81$\pm$0.02 \\
        JO201 & S &   0.0557 &    10.79$\pm$0.06 &     7.12$^{+0.55}_{-0.73}$ &     42$\pm$8 &    176$\pm$7 &      \dots &      \dots &     9.16$\pm$0.04 &     9.09$\pm$0.09 &     8.79$\pm$0.03 \\
    \enddata
\end{deluxetable*}
\addtocounter{table}{-1}
\begin{deluxetable*}{rrrrrrrrrrrr}
\tabletypesize{\footnotesize}
\tablecaption{\it (continued)}
\tablehead{\colhead{ID} & \colhead{sample} &\colhead{redshift} & \colhead{$\log\,(M_\star/M_\odot)$} & \colhead{$R_{\rm e}$} & \colhead{$i$} & \colhead{$P\!A$} & \colhead{$\langle{\rm \frac{O}{H}}\rangle_{<0.1\,R_{\rm e}}$} & \colhead{$\langle{\rm \frac{O}{H}}\rangle_{<0.5\,R_{\rm e}}$} & \colhead{$\langle{\rm \frac{O}{H}}\rangle_{@\,R_{\rm e}}$} & \colhead{$\langle{\rm \frac{O}{H}}\rangle_{\rm disk}$} & \colhead{$\langle{\rm \frac{O}{H}}\rangle_{@\,R_{\rm e}}$CU17} \\
 & & & \colhead{(dex)} & \colhead{(arcsec)} & \colhead{(degree)} & \colhead{(degree)} & \colhead{(dex)} & \colhead{(dex)} & \colhead{(dex)} & \colhead{(dex)} & \colhead{(dex)}}
\decimals
\startdata
        JO204 & S &   0.0450 &    10.61$\pm$0.06 &     5.16$^{+0.25}_{-0.29}$ &     72$\pm$1 &    149$\pm$3 &      \dots &      \dots &     9.14$\pm$0.03 &     9.08$\pm$0.06 &     8.78$\pm$0.02 \\
        JO206 & S &   0.0489 &    10.96$\pm$0.04 &     9.39$^{+1.13}_{-1.20}$ &     64$\pm$3 &    118$\pm$3 &      \dots &      \dots &     8.96$\pm$0.07 &     8.92$\pm$0.11 &     8.68$\pm$0.06 \\
        JO28 & S &   0.0533 &     9.36$\pm$0.08 &     5.40$^{+0.62}_{-0.65}$ &     66$\pm$4 &     32$\pm$5 &     8.58$\pm$0.01 &     8.50$\pm$0.05 &     8.41$\pm$0.11 &     8.38$\pm$0.12 &     8.54$\pm$0.06 \\
        JO45 & S &   0.0452 &     9.16$\pm$0.10 &     3.80$^{+0.23}_{-0.25}$ &     57$\pm$2 &     99$\pm$4 &     8.56$\pm$0.01 &     8.54$\pm$0.04 &     8.50$\pm$0.10 &     8.50$\pm$0.15 &     8.56$\pm$0.03 \\
        JO47 & S &   0.0452 &     9.60$\pm$0.07 &     6.10$^{+0.26}_{-0.39}$ &     45$\pm$4 &     81$\pm$4 &     8.74$\pm$0.03 &     8.68$\pm$0.07 &     8.58$\pm$0.09 &     8.58$\pm$0.10 &     8.62$\pm$0.04 \\
        JO49 & S &   0.0452 &    10.68$\pm$0.06 &     6.48$^{+0.45}_{-0.46}$ &     54$\pm$2 &    117$\pm$5 &      \dots &     \dots &     9.17$\pm$0.04 &     9.11$\pm$0.08 &     8.82$\pm$0.02 \\
        JO60 & S &   0.0586 &    10.40$\pm$0.11 &     4.42$^{+0.36}_{-0.43}$ &     70$\pm$4 &     44$\pm$2 &      \dots &     8.87$\pm$0.05 &     8.70$\pm$0.05 &     8.50$\pm$0.15 &     8.61$\pm$0.02 \\
        JO70 & S &   0.0579 &    10.46$\pm$0.09 &     3.94$^{+0.62}_{-0.60}$ &     40$\pm$6 &     24$\pm$12 &     9.23$\pm$0.00 &     9.17$\pm$0.03 &     9.03$\pm$0.03 &     8.72$\pm$0.23 &     8.73$\pm$0.03 \\
        JO85 & S &   0.0422 &    10.67$\pm$0.09 &     7.04$^{+0.99}_{-1.02}$ &     25$\pm$4 &    158$\pm$6 &      \dots &     \dots &     9.14$\pm$0.02 &     8.92$\pm$0.15 &     8.79$\pm$0.02 \\
        JO93 & S &   0.0419 &    10.54$\pm$0.07 &     9.36$^{+1.33}_{-1.22}$ &     25$\pm$7 &    145$\pm$35 &     9.23$\pm$0.01 &     9.20$\pm$0.03 &     9.06$\pm$0.05 &     8.92$\pm$0.20 &     8.78$\pm$0.03 \\
        JO95 & S &   0.0400 &     9.37$\pm$0.08 &      4.33 &      0 &      0 &     8.47$\pm$0.01 &     8.38$\pm$0.07 &     8.33$\pm$0.13 &     8.28$\pm$0.14 &     8.52$\pm$0.06 \\
        JW100 & S &   0.0551 &    11.47$\pm$0.10 &     6.90$^{+0.40}_{-0.69}$ &     75$\pm$1 &      0$\pm$3 &      \dots &      \dots &     9.24$\pm$0.01 &     9.21$\pm$0.04 &     \dots \\
        JW115 & S &   0.0680 &     9.72$\pm$0.11 &     4.18$^{+0.60}_{-0.64}$ &     69$\pm$3 &     99$\pm$6 &     8.54$\pm$0.02 &     8.50$\pm$0.04 &     8.37$\pm$0.08 &     8.38$\pm$0.10 &     8.51$\pm$0.04 \\
        JW39 & S &   0.0634 &    11.22$\pm$0.08 &     8.33$^{+1.04}_{-1.10}$ &     53$\pm$1 &    100$\pm$4 &      \dots &      \dots &     9.12$\pm$0.06 &     9.09$\pm$0.07 &     8.79$\pm$0.03 \\
        JW56 & S &   0.0461 &     9.05$\pm$0.10 &     2.22$^{+0.18}_{-0.18}$ &     65$\pm$1 &     90$\pm$2 &     8.76$\pm$0.00 &     8.75$\pm$0.01 &     8.77$\pm$0.05 &     8.67$\pm$0.15 &     8.64$\pm$0.04 \\
\enddata
\end{deluxetable*}

\subsection{Gas-phase metallicity}\label{ssec:met}

We compute the gas-phase metallicity, here referring to the oxygen abundance, separately adopting two different metallicity calibrators, the {\sc pyqz} code \citep{dopita2013,vogt2015} and the \citet{curti2017} empirical calibrator based on the O3N2 indicator, to explore their impact on the results.

We use a modified version of {\sc pyqz}~v0.8.2 (F.~Vogt, private communication) that relies on a set of line-ratio grids computed with {\sc mappings\,iv} \citep{sutherland1993,dopita2013}. This version is tested in the range $7.39\le\mathrm{12+\log\,(O/H)}\le9.39$ and adopts the solar oxygen abundance $\mathrm{12+\log\,(O/H)_\odot}=8.69$. The code simultaneously estimates the gas-phase metallicity $\mathrm{12+\log\,(O/H)}$ and the ionization parameter\footnote{The ionization parameter is expressed as $q=U\,c$, where $U$ is the ratio between the density per unit volume of ionizing photons and the gas particle number density, and $c$ is the speed of light.} $\log\,q$, given a set of emission-line ratios. To compute the gas-phase metallicity from the spectra we consider the model grid projected on the line-ratio plane [{\sc O\,iii}]$\lambda$5007/[{\sc S\,ii}]$\lambda\lambda$6717,6731 vs.\ [{\sc N\,ii}]$\lambda$6583/[{\sc S\,ii}]$\lambda\lambda$6717,6731, that, as demonstrated in \citet{dopita2013}, does not present degeneration between the gas metallicity and the ionization parameter and allows an excellent separation of these quantities in the tested range of values. Given the pair of line ratios required by the adopted grid, {\sc pyqz} returns a determined pair of $\mathrm{12+\log\,(O/H)}$ and $\mathrm{\log\,(q)}$. Since the photoionization models cannot fully reproduce all observed line ratios, their predicted fluxes present an uncertainty of $\sim$0.1~dex \citep{kewley2008,dopita2013,blanc2015,mingozzi2020}. In order to investigate the effect of this uncertainty on the gas metallicity error we select 81 line-ratio pairs covering homogeneously the model grid used by our {\sc pyqz} version. For each of the 81 original points we create a sub-set of 1000 line-ratio values randomly distributed around the original point in a normal distribution with a sigma of 0.1~dex. Using {\sc pyqz} we translate each sub-set in a gas metallicity distribution and we calculate their dispersion, that corresponds to the systematic error to associate to the gas metallicity of the original point. To summarize, we find that an uncertainty of 0.1~dex on the models is translated into a systematic error on the metallicity estimate of $\sim$0.05~dex for the highest metallicities (i.e. $\mathrm{12+\log\,(O/H)}=9.39$), up to 0.3~dex for the lowest metallicities (i.e. $\mathrm{12+\log\,(O/H)}=7.39$). In our sample, most of oxygen abundances  are above 8.0, for which the aforementioned systematic uncertainty is smaller than 0.15~dex. Overall, this systematic error is usually dominant with respect to the uncertainty introduced by the errors on the emission line flux measurements.

The calibration of \citet{curti2017} is based on the indicator O3N2, defined as \begin{equation}\label{eq:o3n2}
\mathrm{O3N2=\log\,\biggl(\frac{[\text{\sc O\,iii}]5007}{H\beta}\biggr)-\log\,\biggl(\frac{[\text{\sc N\,ii}]6583}{H\alpha}\biggr)}
\end{equation} that is inversely proportional to the oxygen abundance. It is obtained on stacked spectra of local galaxies in the SDSS-DR7 (Eq.~5 with parameter values in Tab.~2 in \citealt{curti2017}). Their  galaxies have a scatter of 0.21~dex with respect to the derived calibration, with a mean dispersion along the metallicity direction of 0.09~dex. The inferred metallicities are normalized to the solar value $\mathrm{12+\log\,(O/H)_\odot}=8.69$ and the relation is tested within the range $7.6\le\mathrm{12+\log\,(O/H)}\le8.85$. 

We note that we are not taking into account the contamination of the diffuse ionized gas (DIG, see \citealt{haffner2009} for a complete review) and no separation criterion is adopted to exclude the spaxels dominated by the emission of this component, which in principle could have different line ratios \citep[e.g.][]{zhang2017}. 
The DIG properties and its spatial distribution  will be indeed discussed in detail in Tomi\v{c}i\'c et al. (in prep.).

We measure the gas-phase metallicity in each spaxel whose ionized gas flux is powered by SF, requiring a signal-to-noise ratio ($S/N$) $\ge 3$ for all the involved emission lines.\footnote{\cite{mannucci2010} reported that applying high  $S/N$ threshold to the involved emission lines might bias the metallicity measurements. In fact, at the lowest metallicities the flux of the [{\sc N\,ii}]6583 line becomes faint and some spaxels might be removed, shifting high the measured metallicity. Nonetheless, checking the $S/N$ distribution in our galaxies, we find that only the most external spaxels might be affected, but these will be anyways disregarded by the following analysis.}

\subsection{Gas metallicity average of the galaxies}\label{ssec:meanmet}

Thanks to the wide FoV of MUSE (see Sect.~\ref{sec:sample}), we can sample the whole extension of the ionized gas that, excluding the gas tails, reaches a median of $\sim\!3\,R_{\rm e}$ \citep{vulcani2019a}. Taking into account only the star-forming spaxels for which we can infer the gas metallicity, we are able to analyze the chemical properties of the gas up to 1.9  $R_{\rm e}$ for all the reference galaxies, and beyond 2.5  $R_{\rm e}$ for half of them. For the stripping galaxies we can derive the gas-phase metallicity from the center to the gas tails, but in this work we limit our analysis only to the gas within the galaxy stellar disk.

To choose a suitable representative gas-phase metallicity of the galaxy, we explore four possible definitions computing the mean oxygen abundance of the star-forming spaxels in four different radial ranges:

$\langle{\rm O/H}\rangle_{<0.1\,R_{\rm e}}$: spaxels within $0.1\, R_{\rm e}$;

$\langle{\rm O/H}\rangle_{<0.5\,R_{\rm e}}$: spaxels within $0.5\, R_{\rm e}$;

$\langle{\rm O/H}\rangle_{@\,R_{\rm e}}$: spaxels in the range $0.95-1.05\,R_{\rm e}$;

{$\langle{\rm O/H}\rangle_{\rm disk}$}: spaxels beyond $0.5\, R_{\rm e}$ and within the galaxy disk, assuming the same galaxy boundary adopted to derive the total stellar mass.

We decide to exploit the spatially resolved data and not to infer the metallicity from integrated spectra inside the radial ranges of interest to avoid summing together spaxel spectra that in principle have different physical properties (e.g. gas metallicity, ionization parameter, SFR), and thus different emission line ratios. The use of an integrated spectrum entails a luminosity-weighted mean metallicity, while we aim at estimating the average metallicity assigning the same weight to every spaxel. In addition, as discussed by \citet{sanders2017}, gas-phase metallicities derived from global galaxy spectra will be systematically biased by the effects of flux weighting of multiple {\sc H\,ii} regions.

For some galaxies the number of available spaxels is quite low, either given the paucity of ionized gas, or due to the presence of regions not powered by stellar photoionization or due to spaxels with $S/N<3$. To ensure a suitable statistics to calculate the global metallicity, we require that either more than one third of the spaxels, or at least 150 spaxels within the considered radial range have measured oxygen abundance values. If neither condition is fulfilled in the computation of the mean metallicity, the value is discarded. 

The uncertainty that we associate to the mean metallicity corresponds to the standard deviation of the star-forming spaxel metallicities considered in each radial range. For sake of clarity, we do not include the additional error due to the systematic error of the calibration. We note that this must be divided by the root of the number of valid spaxels in the radial bin of interest, giving a contribution on the error budget lower than 0.02~dex.

The metallicity estimates for all our galaxies are listed in Tab.~\ref{tab:controlsample}.

17 galaxies do not fulfill the condition on the minimum number of spaxels with measured oxygen abundance to estimate the $\langle{\rm O/H}\rangle_{<0.1\,R_{\rm e}}$, as they have AGN, LINERs and central composite regions or central gas holes. 
For 10 of them
we also can not derive $\langle{\rm O/H}\rangle_{<0.5\,R_{\rm e}}$ for the same reason and for the remaining seven the $\langle{\rm O/H}\rangle_{<0.5\,R_{\rm e}}$ measure might be underestimated due to the absence of the innermost spaxels that presumably could have higher metallicity.
Instead, the $\langle{\rm O/H}\rangle_{@\,R_{\rm e}}$ and $\langle{\rm O/H}\rangle_{\rm disk}$ measurements
for these galaxies are 
independent of the presence of a central LINER or AGN, as central pixels are not considered in their computation, by definition. Moreover, adopting the \citet{kauffmann2003} separation line to select the star-forming spaxels, instead of the less restrictive separation proposed by \citet{kewley2001}, we make sure to include in the analysis only the spectra with the least or no AGN/LINERs contamination. However, the $\langle{\rm O/H}\rangle_{@\,R_{\rm e}}$ and $\langle{\rm O/H}\rangle_{\rm disk}$ quantities could still suffer the possible absence of the pixels not powered by stellar photoionization or lacking gas along the disk.

We can estimate the $\langle{\rm O/H}\rangle_{@\,R_{\rm e}}$ and $\langle{\rm O/H}\rangle_{\rm disk}$ for  all galaxies but JO135 and JW100, which do not satisfy the minimum number of required spaxels to compute the $\langle{\rm O/H}\rangle_{@\,R_{\rm e}}$; the former when using both calibrators and the latter when using that based on the O3N2 index.

Finally, we report that selecting the star-forming spaxels according to the {\sc O\,i}-based BPT-diagram [{\sc O\,iii}]5007/H$\beta$ vs [{\sc O\,i}]6300/H$\alpha$, instead of the {\sc N\,ii}-based BPT-diagram, does not influence the inferred global galaxy metallicities. The differences between the oxygen abundance at $R_{\rm e}$ derived using the {\sc O\,i} and the {\sc N\,ii} diagrams are at $<$0.1$\sigma$ level on average using both calibrators.
For this reason, in what follows we only show the results based on the star-forming spaxels selected with the {\sc N\,ii}-BPT-diagram.

\subsection{Comparison of the gas metallicities}\label{ssec:zcomp}

In this section we first compare the values of gas-phase metallicity obtained using the radial ranges defined above, with the aim of understanding which definition is the most suitable for our analysis, and then the values of gas-phase metallicity obtained using the different calibrators.

\begin{figure}
\centering
\includegraphics[width=\columnwidth]{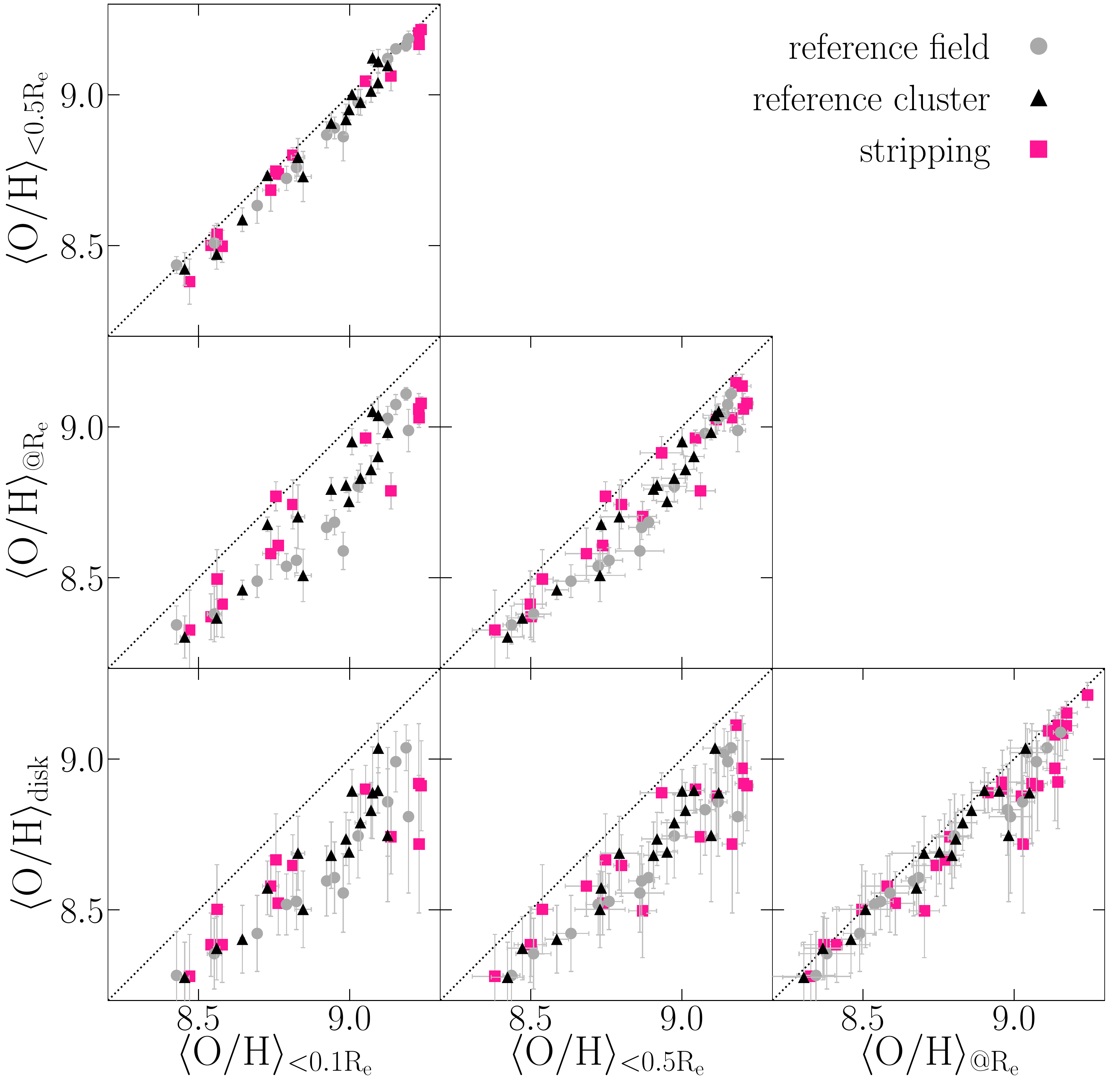}
\caption{Comparison between the gas-phase metallicities inferred with {\sc pyqz} in different radial regions of the galaxies. The gray circles, black triangles and pink squares indicate the reference field, reference cluster and stripping galaxies, respectively. The black dotted lines are the 1:1 relations.\label{fig:zcomp}}
\end{figure}

In Fig.~\ref{fig:zcomp} we compare the four definitions of galaxy gas-phase metallicity inferred with {\sc pyqz}. In each panel, we only plot the galaxies that fulfill the minimum number of required spaxels in the metallicity measurement. The tightest correlation occurs between $\langle{\rm O/H}\rangle_{<0.1\,R_{\rm e}}$ and $\langle{\rm O/H}\rangle_{<0.5\,R_{\rm e}}$ with an rms (root-mean-square) of 0.03~dex. 

We remind that the uncertainty associated to the mean metallicity is the standard deviation of the star-forming pixel metallicities within the considered radial range.

The $\langle{\rm O/H}\rangle_{<0.1\,R_{\rm e}}$ shows the smallest uncertainties because the central pixels have almost the same metallicity. Instead, the $\langle{\rm O/H}\rangle_{\rm disk}$ quantity has the largest uncertainties as the pixel gas metallicities inside the galaxy disk spread on a large range of values.

In the middle panels we show the comparison of $\langle{\rm O/H}\rangle_{<0.1\,R_{\rm e}}$ (on the left), and $\langle{\rm O/H}\rangle_{<0.5\,R_{\rm e}}$ (on the right) with the oxygen abundance at   $R_{\rm e}$. $\langle{\rm O/H}\rangle_{@\,R_{\rm e}}$ exhibits a well-defined correlation with both quantities, in particular with $\langle{\rm O/H}\rangle_{<0.5\,R_{\rm e}}$. The first two bottom panels depict the relations of the central metallicities with $\langle{\rm O/H}\rangle_{\rm disk}$. We still observe a good correlation, although the scatter is slightly larger.

The bottom-right panel in Fig.~\ref{fig:zcomp} illustrates the distribution of $\langle{\rm O/H}\rangle_{\rm disk}$ versus $\langle{\rm O/H}\rangle_{@\,R_{\rm e}}$. Both quantities show a tight correlation, with an rms of 0.07~dex, even if in the high metallicity regime the spread becomes larger. The $\langle{\rm O/H}\rangle_{@\,R_{\rm e}}$ values exhibit smaller uncertainties than $\langle{\rm O/H}\rangle_{\rm disk}$ because the metallicity distribution of the pixels at  $R_{\rm e}$ is narrower than that of the pixels along the whole disk.
This is in agreement with the considerations of \citet{sanchez2013} and \citet{sanchez2017} who compared the total gas-phase metallicity and that at  $R_{\rm e}$ for hundreds of galaxies from the CALIFA survey using different calibrators, including {\sc pyqz}.

It is worth noting that especially $\langle{\rm O/H}\rangle_{\rm disk}$ could be affected by the stripping history for the stripping galaxies and the distribution of the regions not powered by stellar photoionization. In fact, the $\langle{\rm O/H}\rangle_{\rm disk}$ could become less representative when the gas-phase metallicities of excluded regions and that of the stripped gas highly differ from the regions included.

We repeat this analysis considering the oxygen abundance derived with the calibrator based on the O3N2 index. We obtain again a good correlation between the quantities and find that the conclusions discussed for the values derived with the {\sc pyqz} code are also valid for the O3N2 calibrator.

Since one of these methods is sufficient to describe the metallicity of these galaxies, we decide to adopt $\langle{\rm O/H}\rangle_{@\,R_{\rm e}}$ as reference of the galaxy gas-phase metallicity, that allows us to characterize nearly all galaxies avoiding the critical central regions.

\begin{figure}
\centering
\includegraphics[width=0.9\columnwidth]{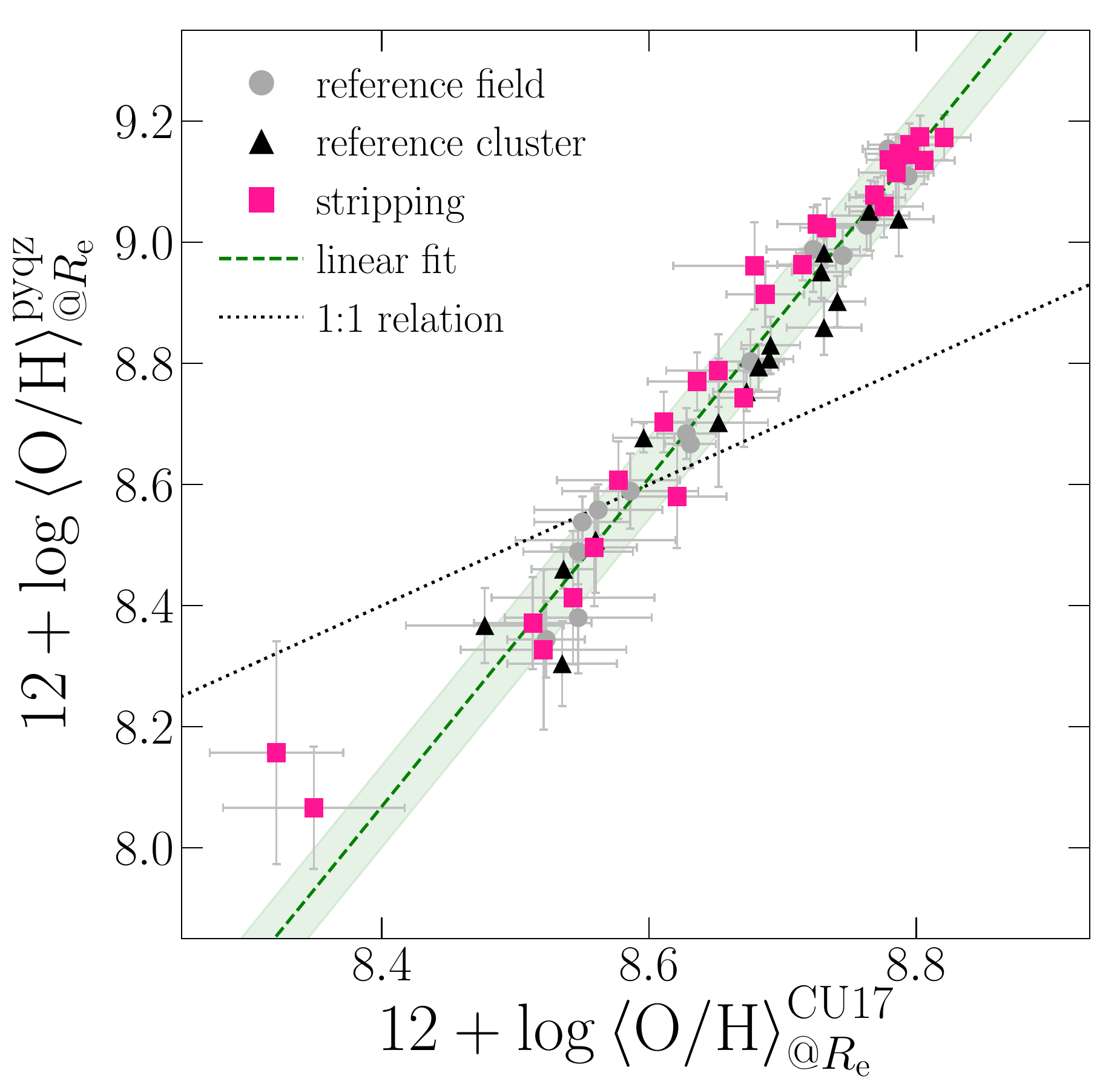}
\caption{Comparison of the gas-phase metallicity at the $R_{\rm e}$ derived using the calibration proposed by \citet{curti2017} and the calibration based on the {\sc pyqz} code. The symbols are as in Fig.~\ref{fig:zcomp}. The green line and the shaded area are the best linear fit of Eq.~\ref{eq:cu17vspyqz} and the rms around the relation, respectively. The black dotted line is the 1:1 relation. \label{fig:cu17vspyqz}}
\end{figure}

We now focus the attention on the metallicity indicators.
The choice of the metallicity calibration has a strong effect on the shape of the MZR and systematic discrepancies are common between the results of different measurement methods \citep{kewley2008}.
Fig.~\ref{fig:cu17vspyqz} illustrates the comparison of the values $\langle{\rm O/H}\rangle_{@\,R_{\rm e}}$ derived with the calibration proposed by \citet{curti2017} for the O3N2 index and the calibration based on the {\sc pyqz} code. We fit the values with a linear relation weighted on the errors of both quantities. The best fit is
\begin{equation}\label{eq:cu17vspyqz}
\begin{split}
\mathrm{\langle Z \rangle}_{\rm pyqz} = & (2.71\pm0.07)\,(\mathrm{\langle Z \rangle}_{\rm CU17}-8.69)\\
& +(8.85\pm0.01),
\end{split}
\end{equation}
where $\mathrm{\langle Z \rangle}=\mathrm{12+\log\langle O/H \rangle}_{@R_{\rm e}}$.

The rms around the relation is 0.07~dex and the Pearson correlation is $r[d.f=57]=0.98$ ($p\sim0$).
For the more metal poor galaxies (JO149 and JO181), we observe the largest differences even though consistent with the linear relation within $1\sigma$. We ascribe this offset to the strong dependence on the ionization parameter of the O3N2 index \citep[see e.g.][]{kruler2017,rodriguezbaras2019}. For high ionization the O3N2 is overestimated, driving to the underestimation of the gas-metallicity using the empirical relation of \citet{curti2017}, in particular for metal-poor gas. By a visual inspection of the spatially-resolved distribution of the $\log\,q$ quantity, derived by the {\sc pyqz} code, we detect that the central regions of both galaxies are dominated by large ionization parameter values. Therefore, the metallicity of JO149 and JO181 obtained by the O3N2 calibrator could be biased towards lower values.

Overall, although the absolute values of the gas metallicity are different, the relative values remain reliable and the correlation is narrow indicating that the metallicity distribution within the sample is independent of the choice of the calibrator. 

In what follows we
show only the results obtained with the {\sc pyqz} calibration.
\section{Results}\label{sec:result}

\subsection{Mass-metallicity relation}\label{ssec:mzr}

\begin{figure}
\centering
\includegraphics[width=\columnwidth]{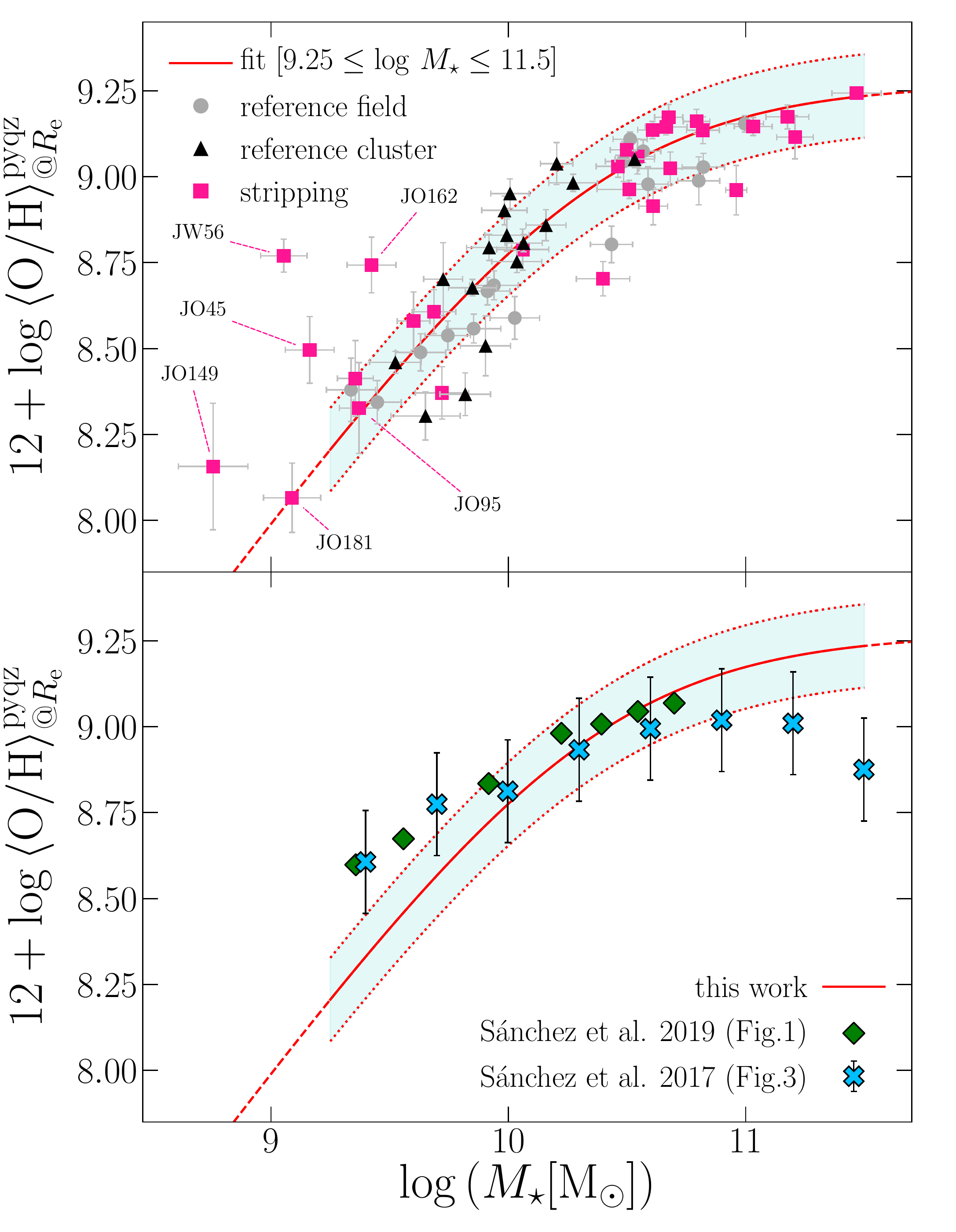}
\caption{{\it Upper panel}: Relation between the galaxy stellar mass and the {\sc pyqz} gas-phase metallicity at $R_{\rm e}$. Symbols are as in Fig.~\ref{fig:zcomp}. The red line represents the best fit assuming Eq.~\ref{eq:mzr}. The shaded area between the red dotted lines is the scatter around the fit. The flagged galaxies are commented in the text. {\it Bottom panel}: Comparison between the MZR derived in this work (red curve  with the shaded area) and the median metallicities at different stellar masses derived from CALIFA \citep[blue crosses,][]{sanchez2017} and SAMI \citep[green diamonds,][]{sanchez2019} samples.}\label{fig:mzrpyqz}
\end{figure}

One of the aims of this work is to analyze the galaxy gas-phase metallicity versus stellar mass diagram, illustrated in the upper panel of Fig.~\ref{fig:mzrpyqz}, for the three galaxy samples. Field and cluster galaxies, as well as reference and stripping galaxies are generally located on a relatively tight correlation, with no apparent signs of offset between the samples, with the exception of four among the least massive stripping galaxies that lie well above the general relation. 

Previous studies have shown that the gas-phase metallicity has a steep dependence on stellar mass for galaxies with $M_\star\leq $10$^{10}$~M$_\odot$ and then it becomes flatter at higher masses \citep[e.g.][]{tremonti2004,kewley2008}.
We fit the data with the function:
\begin{equation}
\label{eq:mzr}
12+\mathrm{\log\,(O/H)} = \widetilde{Z} - \log\, \Biggl\{ \frac{1}{2}\, \biggl[ \biggl(\frac{M_\star}{\widetilde{M}}\biggr)^{-\gamma} +1 \biggr] \Biggr\}.
\end{equation}
In the log-log plane, this relation is linear for masses lower than $\widetilde{M}$ with a slope proportional to the parameter $\gamma$. At $M_\star=\widetilde{M}$ it reaches $\widetilde{Z}$ and flattens out at higher masses.
We restrict the fitting where the relation appears well-defined, so in the stellar mass range $9.25\le\log\,(M_\star/M_\odot)\le11.5$. With this choice,
four low mass galaxies, all stripping galaxies (JO45, JO149, JO181 and JW56), are excluded from the fit.

Table~\ref{tab:fitmzr} lists the best-fit parameters obtained considering the uncertainties on the oxygen abundances. In Fig.~\ref{fig:mzrpyqz} we also show the rms around the relation (0.12~dex) in the considered mass range.\footnote{We note that even excluding the galaxies hosting an AGN the fitted curve does not appreciably change.}

\begin{table}[]
\centering
\caption{Best fit parameters of Eq.~\ref{eq:mzr}.} \label{tab:fitmzr}
\begin{tabular}{crcl}
\hline
\hline
Parameter & Value & Uncertainty & \\
\hline
\decimals
$\widetilde{Z}$ & 8.97 & 0.04  & (dex)   \\
$\log\,\widetilde{M}$ & 10.35 & 0.10  & (dex) \\
$\gamma$ & 0.93 & 0.09   \\
\hline
\end{tabular}
\end{table}

\begin{figure}
\centering
\includegraphics[width=\columnwidth]{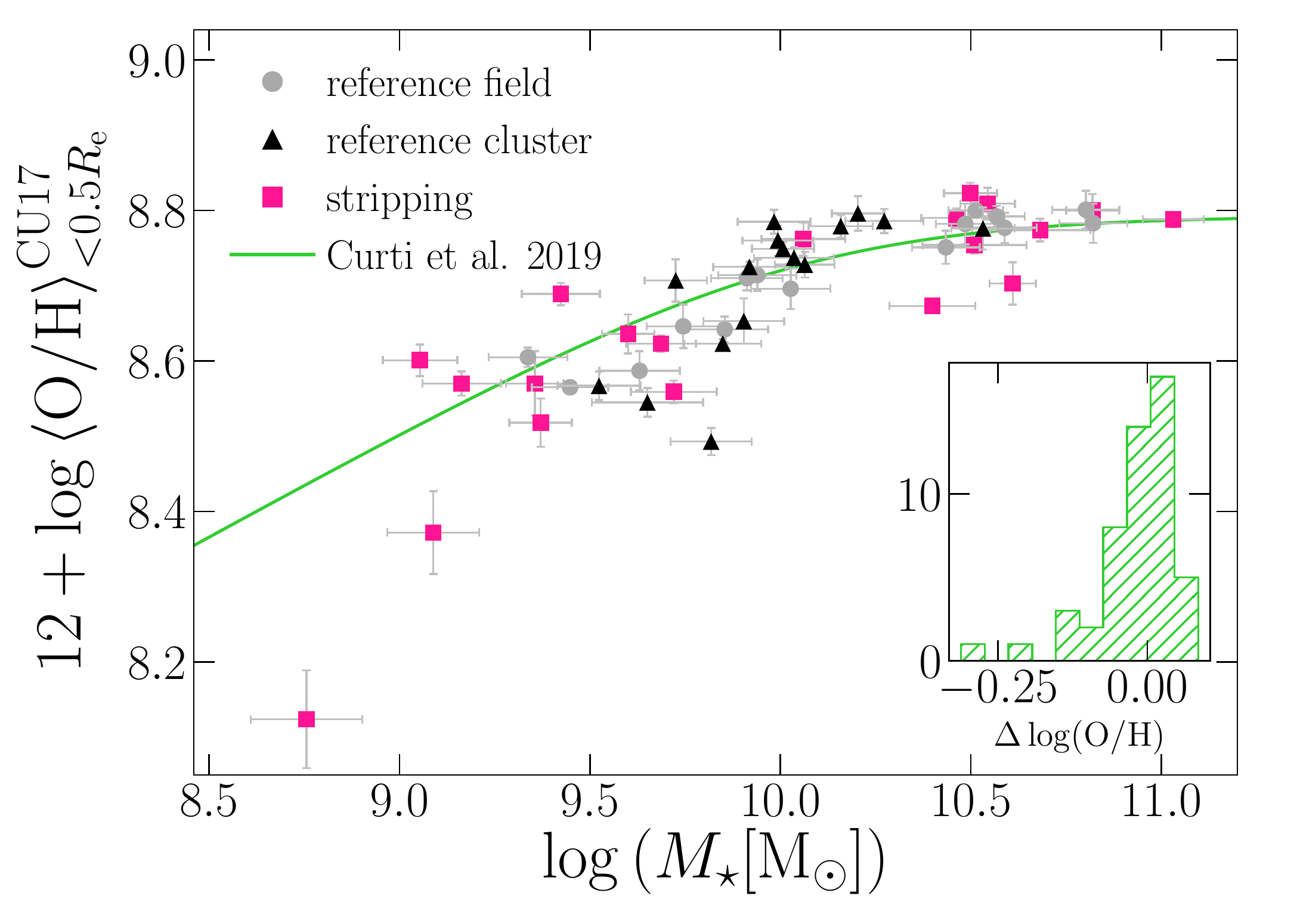}
\caption{Relation between the galaxy stellar mass and the gas-phase metallicity within $0.5\,R_{\rm e}$ based on the O3N2 index. Symbols are as in Fig.~\ref{fig:zcomp}. The green line represents the MZR from \citet{curti2020}. The inner panel shows the histogram of the metallicity residuals of the galaxies with respect to the \citeauthor{curti2017} fit). \label{fig:MZRCU17}}
\end{figure}

In the bottom panel of Fig.~\ref{fig:mzrpyqz} we compare the MZR of this work with the median metallicities at different stellar masses derived by \citet{sanchez2017} and \citet{sanchez2019} using the CALIFA and SAMI samples, respectively, and adopting the {\sc pyqz} code. Although they estimate the metallicities at the effective radius from the linear fitting of the galaxy abundance gradient (not by averaging the metallicity of the spaxels) we observe a good agreement, but at the highest and lowest masses where our MZR deviates.

Recently, \citet{curti2020} fitted the MZR relation for $\sim$150\,000 SDSS galaxies at $z>0.027$ selected from the MPA/JHU catalog, exploiting the integrated emission fluxes inside the SDSS fiber with $3''$ diameter aperture - corresponding to a sampling of 1.6 kpc at $z=0.027$.  \citet{curti2020} estimated the gas-phase metallicity using a combination of the calibrations developed in \citet{curti2017} and a new series analogously determined based on nine different indexes.

We compare their MZR with
the gas-phase metallicity within $0.5\,R_{\rm e}$ derived with the calibrator of \citet{curti2017} based on the O3N2 index.\footnote{To be more consistent, we should use the  $\langle{\rm O/H}\rangle_{<0.1\,R_{\rm e}}$, as our radial range is larger than 1.6 kpc (the median $R_e$ of our sample is 4.9~kpc), but as described in Sec 3.3 many galaxies lack the measurement of $\langle{\rm O/H}\rangle_{<0.1\,R_{\rm e}}$. Fig.3 showed that $\langle{\rm O/H}\rangle_{<0.1\,R_{\rm e}}$ and $\langle{\rm O/H}\rangle_{<0.5\,R_{\rm e}}$ well correlate.} Figure~\ref{fig:MZRCU17} shows that, overall, the fit provided by \citet{curti2020} well represents also our data-points, with an rms of the residuals of 0.07 dex. However, at $M_\star<10^{10}\,{\rm M_\odot}$ our data seems to follow a steeper trend than the relation of \citet{curti2020}. Moreover, the overabundance observed for the least massive galaxy JO149 disappears, but it could be due to the bias of the O3N2 calibrator, as discussed in Sect.~\ref{ssec:zcomp}.

\subsection{Analysis of the residuals along the MZR}

\begin{figure}
\centering
\includegraphics[width=\columnwidth]{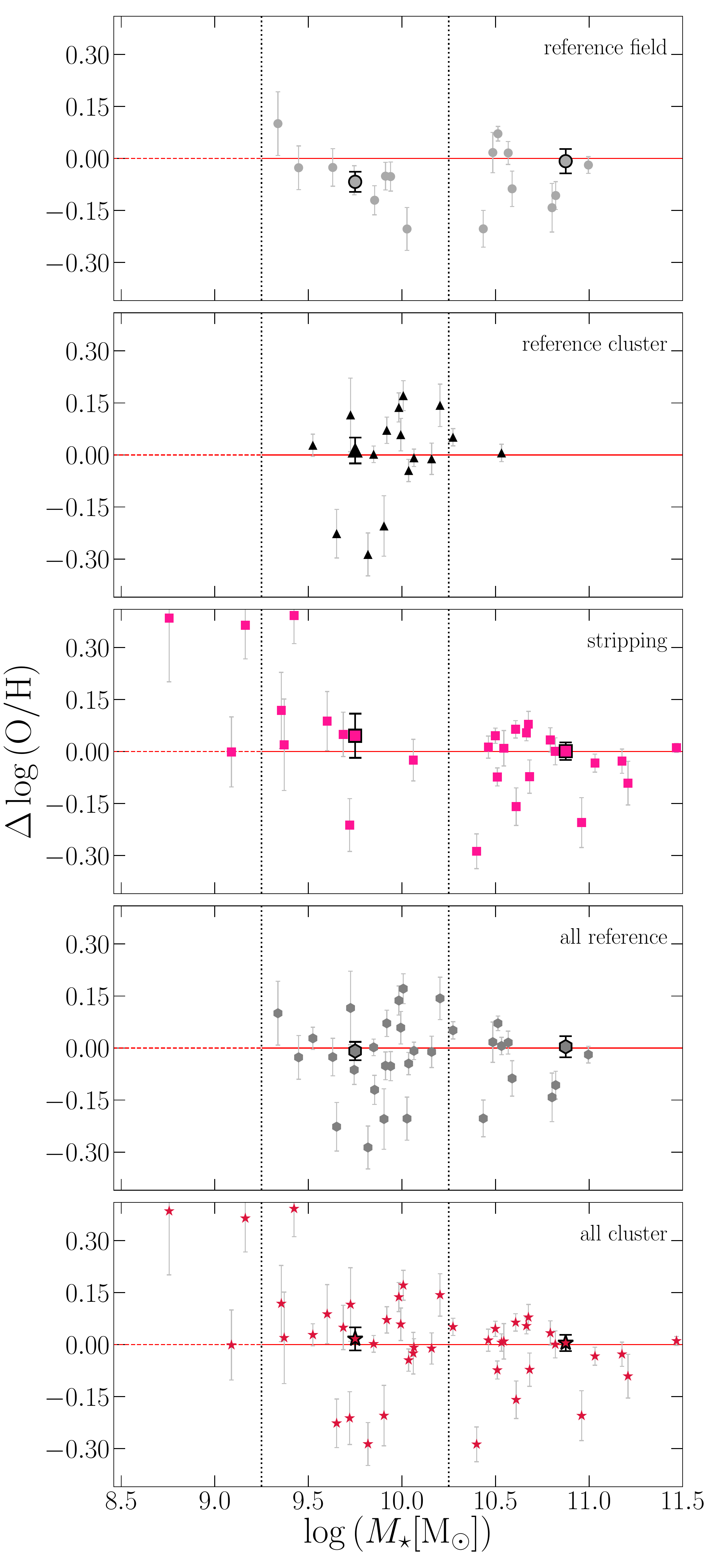}
\caption{From the top to the bottom panel: residuals of the MZR for the reference field (gray circles), reference cluster (black triangles), stripping (pink squares), reference (reference field+cluster; dark gray hexagons) and cluster (reference cluster+stripping; red stars) galaxies. In each panel, the horizontal red line corresponds to the MZR of Fig.~\ref{fig:mzrpyqz} and the vertical dotted black lines are the mass separation lines at $M_\star=10^{9.25}~{\rm M_\odot}$ and $M_\star=10^{10.25}~{\rm M_\odot}$ to divide the samples in three mass bins. In each panel, the bigger black edged symbols are the error weighted average in the corresponding mass bin along with the standard errors. JW56 is not shown as its residual ($\Delta\log({\rm O/H})=0.73$~dex) is out of the panel limits.\label{fig:binnone}}
\end{figure}

To investigate the distributions of the galaxy samples along the MZR, in Fig.~\ref{fig:binnone} we compute the residuals of the {\sc pyqz} gas-phase metallicity ($\Delta\log({\rm O/H})$) from the best fitting relation of Fig.~\ref{fig:mzrpyqz}. From top to bottom, the figure shows the residuals of the reference field, reference cluster, stripping galaxies and the overall samples of the reference and cluster galaxies, respectively. We also divide the total sample (reference$+$stripping galaxies) in two different equally populated stellar mass bins: $10^{9.25}\leq (M_\ast/M_\sun)<10^{10.25}$ containing 29 galaxies and  $10^{10.25}\leq (M_\ast/M_\sun)<10^{11.5}$ containing 28 galaxies. 
We then compute the gas metallicity residuals of the galaxies  with $M_\star<10^{9.25}$~M$_\odot$ extrapolating the MZR at low masses, therefore these values must be taken with caution.

For galaxies in both mass bins separately,
we calculate the error-weighted average value of the residuals along with its standard error. For the reference cluster galaxies we do not compute the mean values in the high mass bin as there are only two objects.

The most relevant outcome that we detect occurs at $10^{9.25}\leq (M_\ast)<10^{10.25}~{\rm M_\odot}$, where all  reference field galaxies but one are located below the relation and the mean value of their residuals is lower than that of the cluster galaxies, both reference and stripping, at $>$1$\sigma$ level.
At high masses the shift disappears and, in all the sub-samples, the mean values of the residuals are consistent with zero.

This result appears in agreement with \citet{maier2019a} that observed an overabundance of gas-phase metallicity in cluster galaxies with $M_\star<10^{10.5}~{\rm M_\odot}$ compared to field galaxies of similar masses.

No strong offset emerges between the reference cluster galaxies and the stripping ones. In addition, for these galaxies the spread of the residuals tends to decrease with increasing stellar mass, suggesting that less massive objects are more sensitive to the processes affecting the chemical evolution.

\section{Discussion}\label{sec:discus}

In the previous section we have shown that field galaxies, cluster galaxies and RPS galaxies 
follow on average a well-defined MZR, but significant 
deviations are also observed at low and intermediate masses between field and cluster galaxies. This means that, although
the stellar mass is the parameter that mainly drives the chemical enrichment in galaxies, other physical conditions can have implications for the gas-phase metallicity evolution.

In this section we investigate the scatter around the MZR and probe possible secondary dependencies.

\subsection{The effects of the ram-pressure stripping}

At masses $>10^{9.25}\,{\rm M_\odot}$, the mean metallicity of stripping galaxies are consistent with those of the reference cluster galaxies, so the gas stripping mechanism apparently does not determine a crucial alteration of the gas-phase metallicity at $R_{\rm e}$. However, the RPS could entail the redistribution of gas, hence of the metals, inside the disk, producing either a reduction or an increase of the gas metallicity. 

On one side, hydrodynamical simulations have found that the RPS can produce a large infall rate of gas to the inner regions, both pushing the gas to the center and by flowing of the galactic gas that lost angular momentum as a consequence of the interaction with the non-rotating ICM \citep{vollmer2001,tonnesen2009,ramosmartinez2018}. If the infalling gas was metal-poor, the central metallicity, and likely that at the $R_{\rm e}$ too, could be lower than the expectation for an undisturbed galaxy with same mass. In addition, the inward gas flow could feed the central galactic black hole and triggers its activity \citep{poggianti2017b}; then, the consequent AGN feedback can enhance the metal ejection from the galaxy \citep{derossi2017,chisholm2018} and lead to both a reduction of the central gas-phase metallicity and an increment in the external regions.

On the other hand, simulations by \cite{schulz2001} have shown that RPS can also displace the gas with respect to the galaxy halo center, bringing the innermost metal-rich gas to larger radii, especially in low-mass galaxies. This possibility represents a critical issue for the $\langle{\rm O/H}\rangle_{@\,R_{\rm e}}$ estimation of the extreme stripping galaxies, whose truncation radius (the lower radius of the removed gas) is equal or less that the $R_{\rm e}$.

This complex scenario could be seen with the residuals of the MZR for the stripping galaxies, plotted in the middle panel of Fig.~\ref{fig:binnone}. At stellar masses greater than $10^{9.7}~{\rm M_\odot}$ there are several stripping galaxies below the fitted MZR, even if their values are consistent with those of the reference sample, while at low masses we observe stripping galaxies with higher gas-phase metallicity with respect to the relation. The Pearson correlation coefficient states that the anti-correlation with galaxy mass for stripping galaxies is significant ($r[d.f.=25]=-0.55$, $p=0.003$). However, the trend is no longer relevant if we exclude the 4 low mass galaxies with the largest residuals above the MZR: JO162 ($\Delta\log({\rm O/H})=0.39$~dex), JO45 ($\Delta\log({\rm O/H})=0.36$~dex), JW56 ($\Delta\log({\rm O/H})=0.73$~dex) and JO149 ($\Delta\log({\rm O/H})=0.39$~dex). To better comprehend if their metallicity differences are due to the gas relocation as a consequence of the stripping, we explore the spatially-resolved distribution of their gas metallicity. 

\begin{figure*}
\centering
\includegraphics[width=0.8\textwidth]{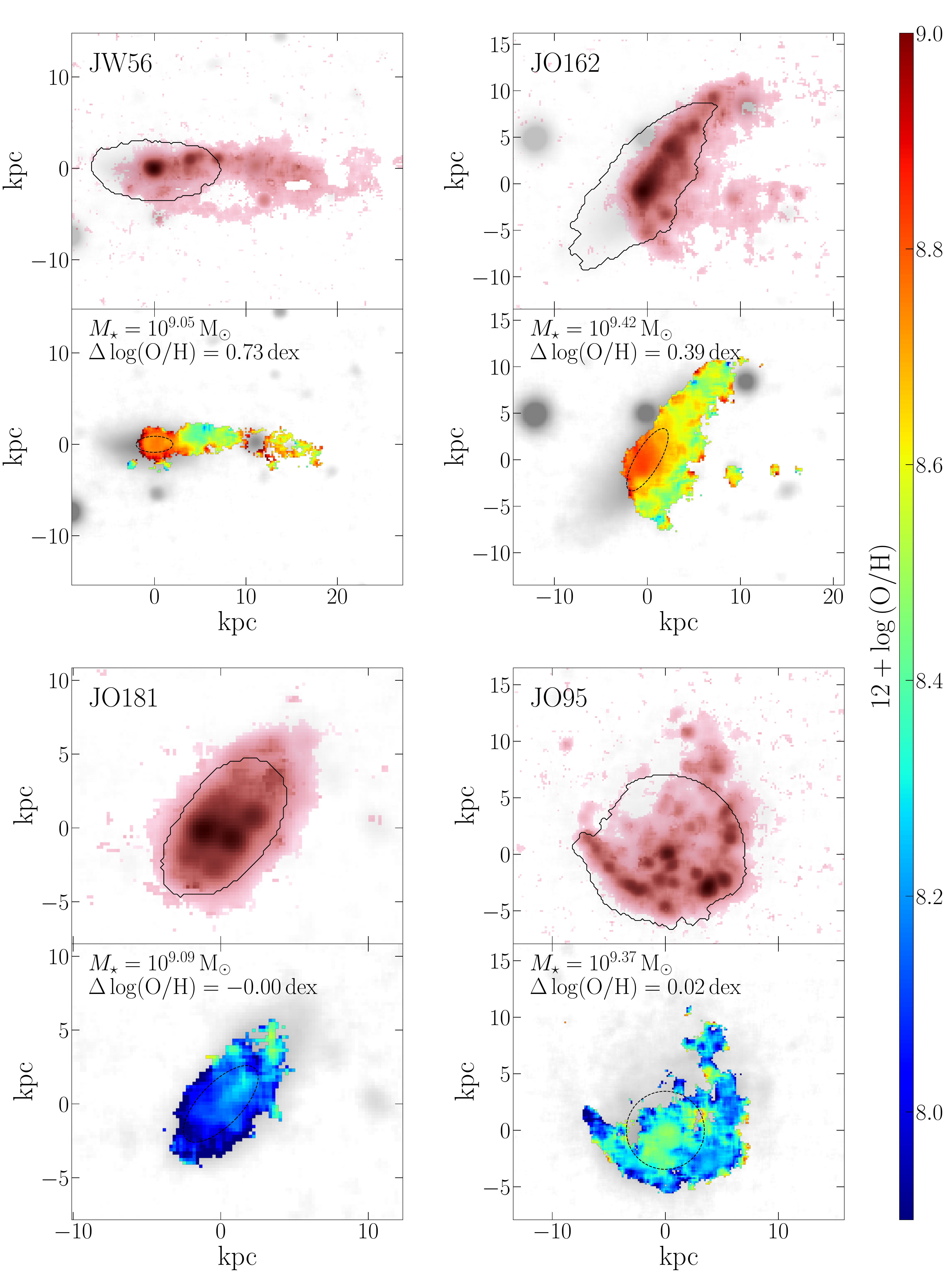}
\caption{H$\alpha$ emission and {\sc pyqz} gas-phase metallicity maps of the galaxies JW56, JO162, JO181 and JO95 (all stripping galaxies) superimposed on the stellar continuum underlying the H$\alpha$ line (gray). JW56 and JO181 have $M_\star\approx 10^{9.07}\,{\rm M_\odot}$; JO162 and JO95 have  $M_\star\approx 10^{9.40}\,{\rm M_\odot}$. For each galaxy: the {\em top panel} shows the H$\alpha$ emission with $S/N>4$ (pink) and the contour of the galaxy disk (black line); the {\em bottom panel} depicts the {\sc pyqz} gas-phase metallicity of the star-forming pixels (according to the color code) and the $R_{\rm e}$ (black dashed ellipse).}
\label{fig:metmaps}
\end{figure*}

Figure~\ref{fig:metmaps} illustrates the H$\alpha$ emission and the gas-phase metallicity maps of the stripping galaxies with the largest metallicity residuals (JW56 and JO162) and those of two stripping galaxies with similar mass, but having $\langle{\rm O/H}\rangle_{@R_{\rm e}}$ consistent with the MZR (JO181 and JO95). In the gas-phase metallicity maps we only plot the star-forming spaxels according to the \citet{kauffmann2003} separation line on the {\sc N\,ii}-based BPT-diagram and superimpose the $R_{\rm e}$ projected on the galaxy disk. 
The metallicity enhancement measured at the $R_{\rm e}$ for JW56 and JO162 is not due to the redistribution of the inner metal-rich gas outwards the external regions, but it depends on an intrinsically overall high chemical abundance. In addition, their high $\langle{\rm O/H}\rangle_{@R_{\rm e}}$ values are not even due to the small size of their effective radii that might sample inner and higher metallicities, indeed these stripping galaxies show chemical overabundances also when the mass-metallicity distribution is explored using the $\langle{\rm O/H}\rangle_{\rm disk}$ estimates. The same is valid for JO149 and JO45.
An hypothesis could be that this intrinsic overabundance of gas-phase metallicity might be due to a fast self-enrichment as a consequence of the SFR enhancement induced by the ram-pressure process. For JO149 and JO162 this hypothesis could be true, in fact these two galaxies have a surplus of SFR ($\Delta\log({\rm SFR})=0.58$ and 0.23~dex, respectively) with respect to the mass-SFR relation inferred by \citet{vulcani2018b} using a control sample from GASP. On the other hand, JW56 and JO45 do not show the SFR boost ($\Delta\log({\rm SFR})=0.01$ and -0.14~dex, respectively), suggesting that for these galaxies
the metallicity residuals could be connected to other physical properties.

Lastly, we have to consider that for some stripping galaxies there are geometric effects connected to the stripping angle. The three-dimensional location of the removed gas, mainly metal-poor, could be such to overlap the galaxies along the line-of-sight and, thus, entails an artificial underestimation of the gas-metallicity in the disk.
To investigate in detail this point of view, the spatially-resolved maps and the radial profiles of the gas-phase metallicity will be analyzed in a future paper (Franchetto et al., in prep.).

\subsection{The dependence on the star formation rate}

It has been shown that at a given mass galaxies with high SFR are characterized by lower gas metallicity than those with low SFR \citep{mannucci2010,laralopez2010,hunt2012,yates2012}. 
We can now probe the behaviour of our galaxies in this framework and verify if their gas-phase metallicity is consistent with the hypothesis of a secondary dependence on the SFR.

\begin{figure}
\centering
\includegraphics[width=\columnwidth]{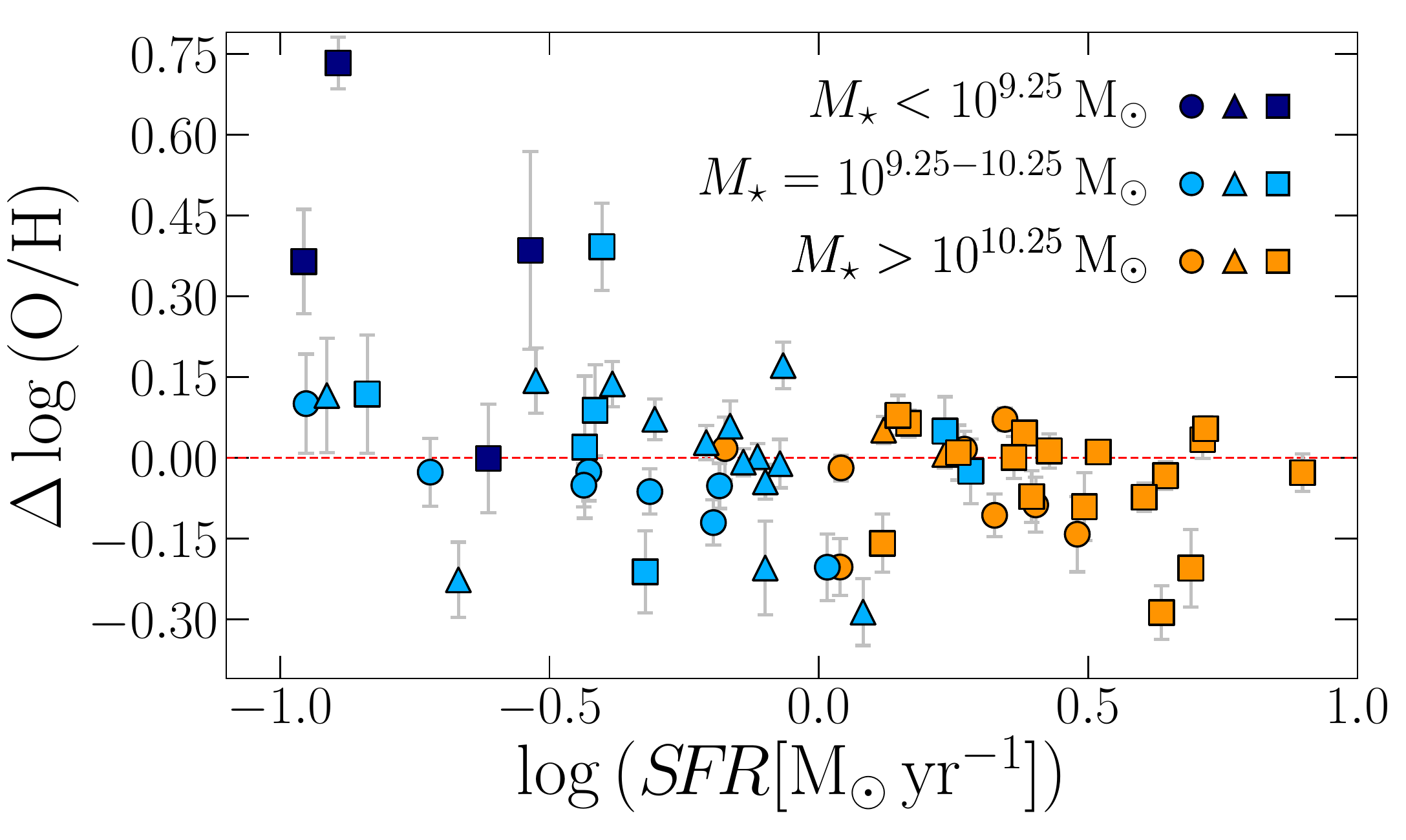}
\caption{Residuals of the MZR versus SFR. Circles, triangles and squares indicate the reference field, reference cluster and stripping galaxies, respectively. Colors refer to the galaxy stellar masses as in the legend. The horizontal red line corresponds to the MZR of Fig.~\ref{fig:mzrpyqz}.}
\label{fig:MZR_SFR}
\end{figure}

Fig.~\ref{fig:MZR_SFR} shows the gas-phase metallicity residuals of the MZR versus the SFR. The SFR values are taken from \citet{vulcani2018b} and Gullieuszik et al.\ (submitted), who compute and sum the SFRs of spaxels inside the galaxy boundary from the H$\alpha$ flux corrected for stellar and dust absorption, excluding those spaxels classified as AGN or LINERS by the \citet{kewley2001} separation line in the {\sc N\,ii}-based BPT-diagram, and adopting \citet{kennicutt1998}'s relation for a \citet{chabrier2003} IMF. Galaxies with negative or close to zero metallicity residuals span almost the entire range of estimated SFR values, while galaxies with the largest positive residuals mainly occupy the region at low SFRs. No galaxies with high SFR are found much above the MZR.
Considering the galaxies with $M\ast>10^{9.25} M_\odot$, we observe a moderate anti-correlation between the quantities, also supported by the Pearson correlation coefficient ($r[d.f.=54]=-0.28$, $p=0.04$). This anti-correlation is even more significant when we also include the galaxies with stellar masses below $10^{9.25}~{\rm M_\odot}$, for which the residuals are obtained by extrapolation of the MZR ($r[d.f.=58]=-0.44$, $p=0.0005$). Galaxies with the largest overabundance of gas-phase metallicity, in particular JW56, have very low total SFRs.

We have also investigated the the gas-phase metallicity residuals of the MZR versus the sSFR (plot not shown), but no statistically significant trend emerged.

\citet{vulcani2018b} have shown that galaxies undergoing RPS have SFR values up to 0.2 dex larger than control sample galaxies of similar mass.
Since this SFR surplus is likely due to the compression of the gas in the disk as a consequence of the impact with the ICM, to suitably compare the SFR of reference and stripping samples we scale down the SFR of the stripping galaxies and calculate again the Pearson correlation coefficient. The anti-correlation becomes more pronounced, with a higher level of significance ($r[d.f.=58]=-0.50$, $p=0.00004$), meaning that the metallicity scaling relation observed for the reference sample are still valid for the stripping galaxies, and the chemical evolution is driven by the same physical processes. 

As discussed in the introduction, several studies have reported the existence of the mass-SFR-metallicity relations (or FMR).
A direct comparison with this relation could provide a further proof of the secondary dependence on the SFR.
Nonetheless, this analysis is not trivial because the shape of the FMR  is connected to the measurement methods of the gas metallicity and SFR, and to the selection criteria of the sample \citep{cresci2019}.

\begin{figure*}
\centering
\includegraphics[width=1\columnwidth]{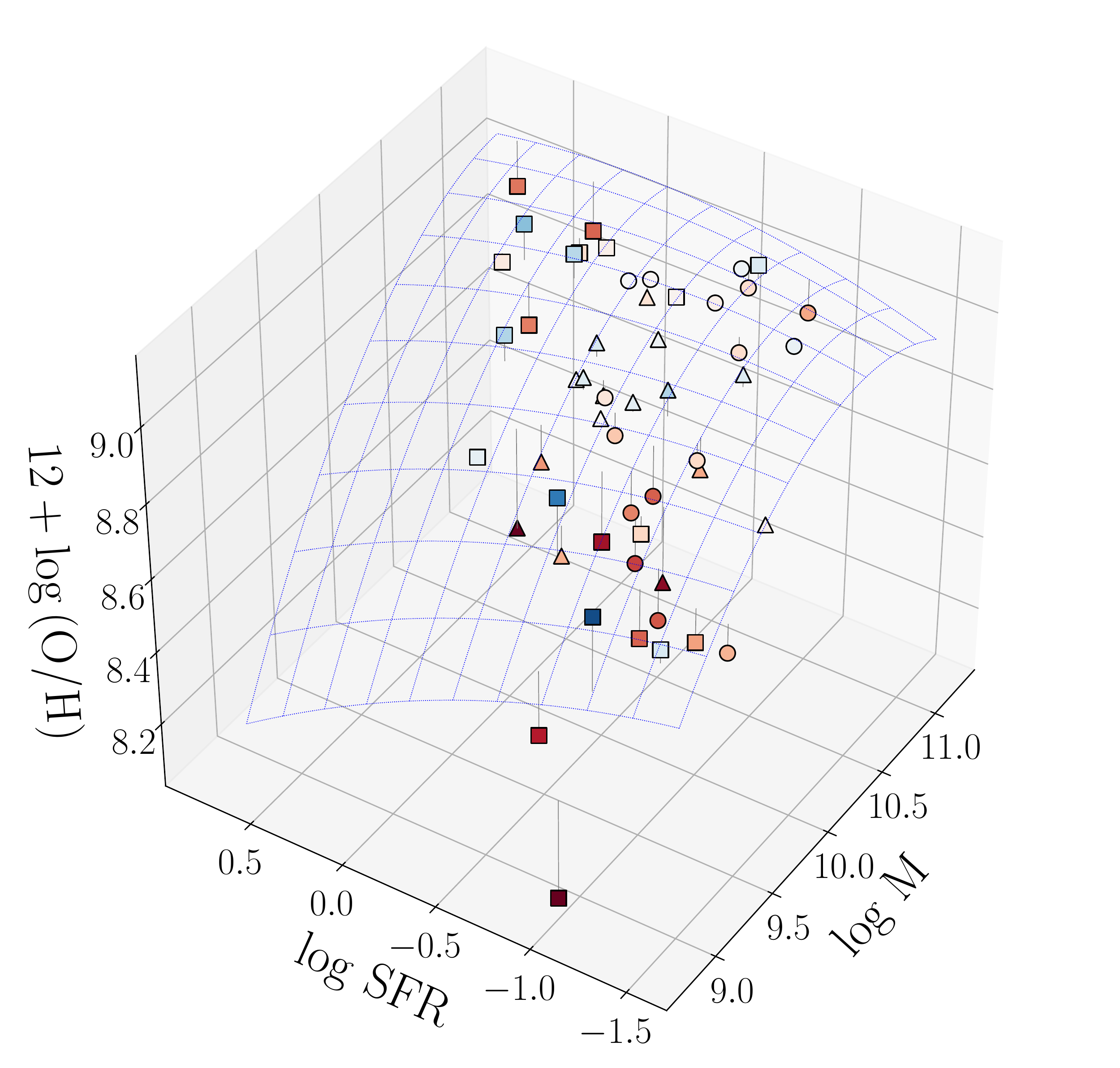}
\includegraphics[width=1\columnwidth]{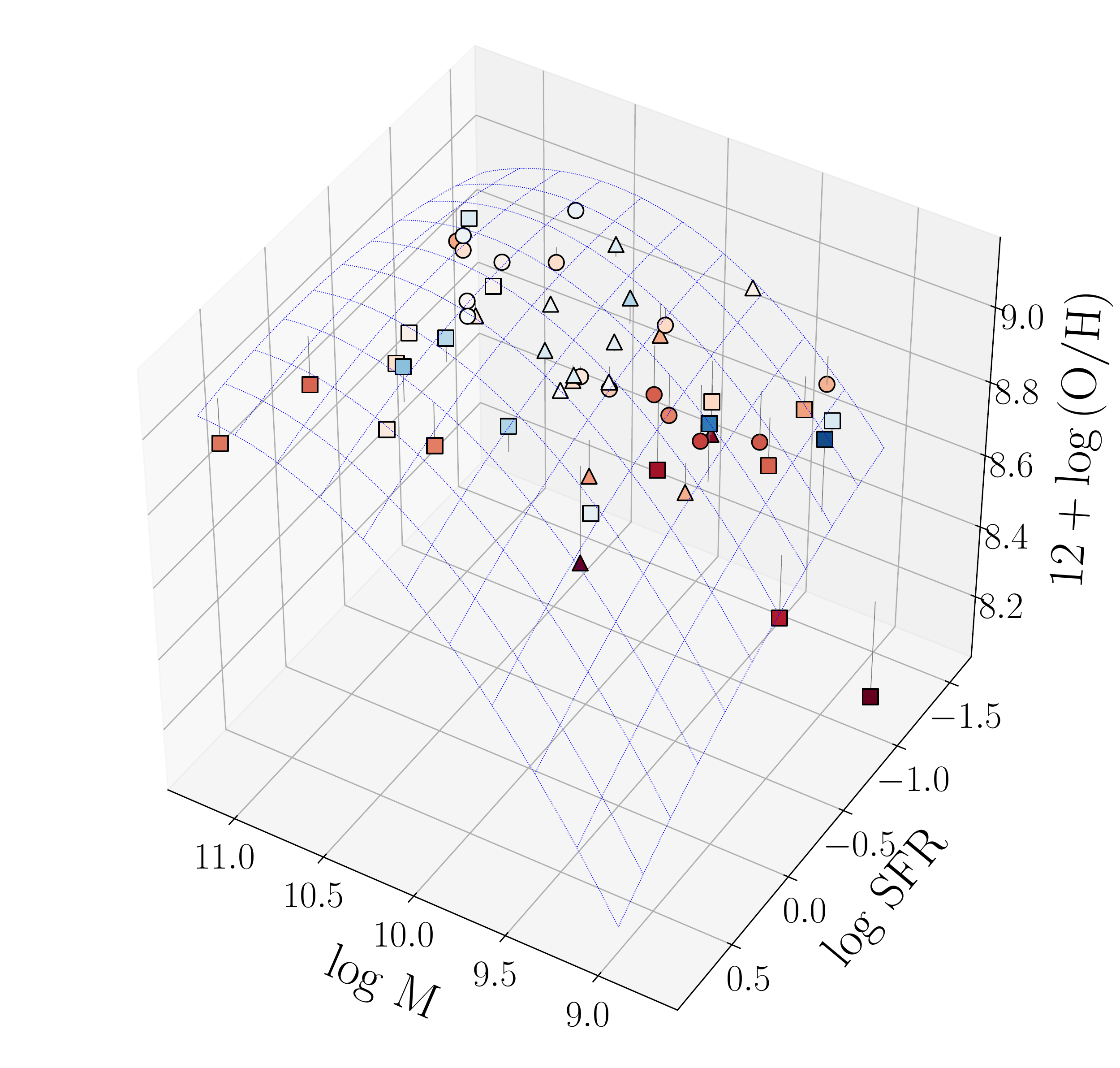}
\includegraphics[width=1.1\columnwidth]{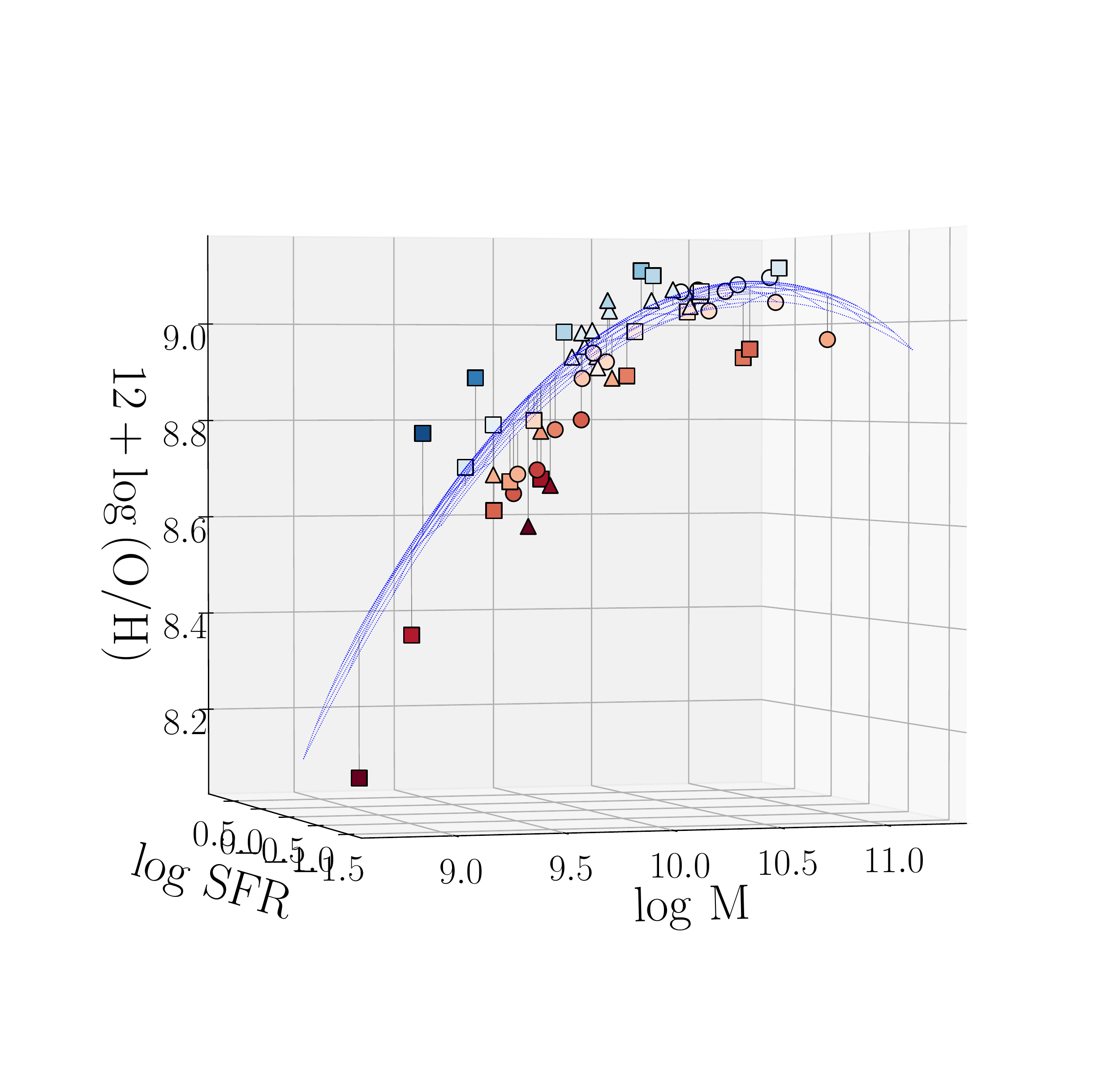}
\caption{Three projections of the M$_\star$-SFR-metallicity space. The surface is the FMR of \citet{mannucci2010}. The circles, triangles and squares correspond to the reference field, reference cluster and stripping galaxies respectively. The colors and the vertical lines indicate the metallicity distance from the surface (blue and red correspond to positive and negative differences, respectively).  \label{fig:FMRfiber}}
\end{figure*}

To probe if our sample is consistent with the FMR derived by \citet{mannucci2010} we attempt a simple emulation of their analysis
and verify the distribution of our datapoints around the surface depicted by their FMR. Briefly, the authors used the SDSS-DR7 dataset of $\sim$140\,000 galaxies with $0.07<z<0.3$. SDSS spectroscopic observations were carried out with fibers of $3''$ aperture, which sample a large fraction of the central galaxy region corresponding to $\sim$4--11~kpc depending on redshift. \citet{mannucci2010} apply a threshold of $S/N>25$ on the H$\alpha$ fluxes and exclude AGN-like galaxies adopting the diagnostics line of \citet{kauffmann2003}. The total stellar masses are taken from the MPA/JHU catalog of the SDSS-DR7 database and scaled from the \citet{kroupa2001} to the \citet{chabrier2003} IMF. The dust-corrected H$\alpha$ flux inside the fiber is converted in SFR using the \citet{kennicutt1998} relation for a \citeauthor{chabrier2003} IMF. No aperture correction is applied. The oxygen abundances are measured adopting the calibrator described in \citet{maiolino2008} for the N2 ([{\sc N\,ii}]6584/H$\alpha$) and R23 (([{\sc O\,ii}]3727+[{\sc O\,iii}]4958,5007)/H$\beta$) indexes and taking the average of the two values.
\citeauthor{mannucci2010} first fit the mass and metallicity values with a polynomial equation in order to derive the MZR of their sample \citep[Eq.~1]{mannucci2010}, then including the SFR values they find the FMR \citep[Eq.~2]{mannucci2010}.

We mimic the emission line dataset of \citet{mannucci2010}, summing, for each galaxy, the flux of all the spaxels inside a circular aperture of variable size, placed in the center of the galaxy, in such a way to sample always a diameter of 5~kpc at the redshift of the galaxies.
No spaxel inside the aperture is masked before the integration. For each galaxy, the $S/N$ of H$\alpha$ integrated flux is always greater than 25. No $S/N$ cut is applied for the other emission lines.
We exclude AGN-like galaxies applying the diagnostic line of \citet{kauffmann2003} and measure the SFR using the \citet{kennicutt1998}'s relation.
Finally, we compute the oxygen abundance employing the calibration of \citet{maiolino2008} for the N2 and R3 ([{\sc O\,iii}]5007/H$\beta$) indexes, and then we take the average value. We use the R3 index to substitute the R23 one because the MUSE data do not cover the [{\sc O\,ii}]3727 line.

\begin{figure*}
\plotone{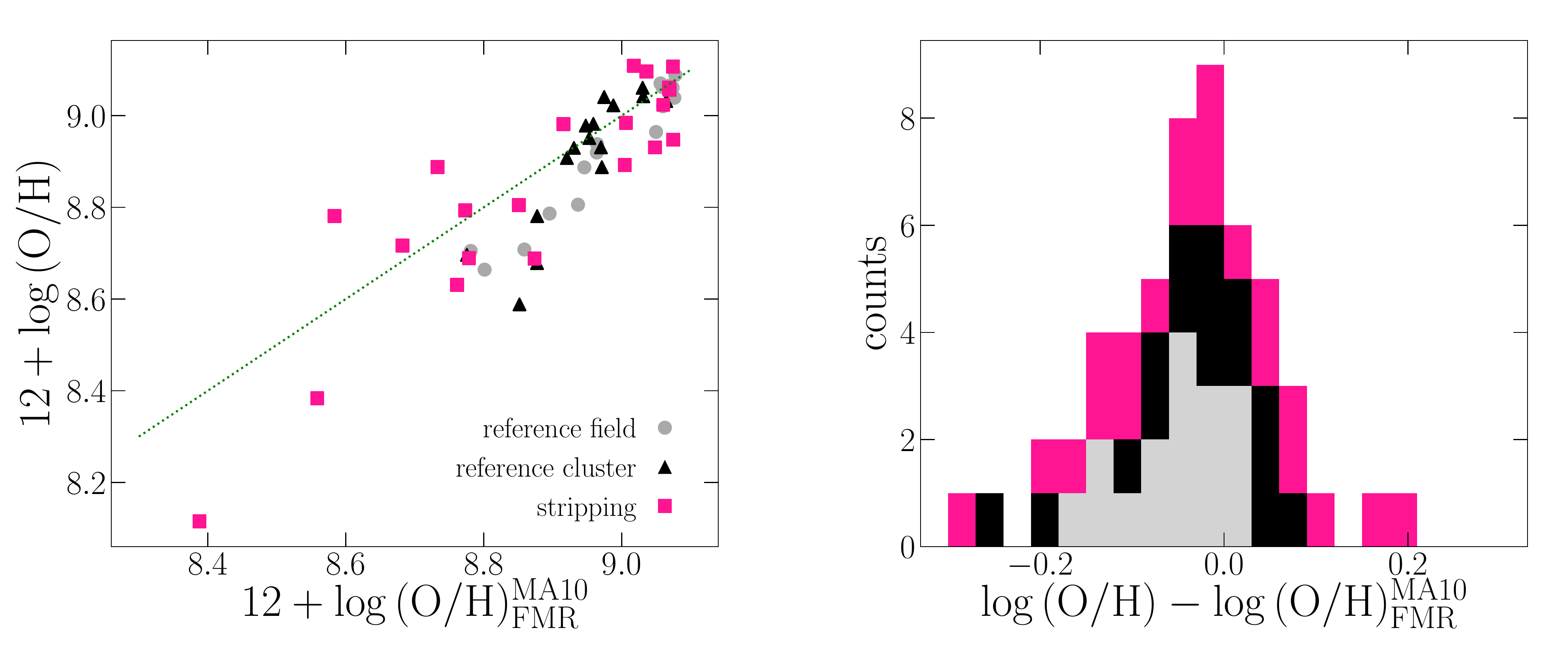}
\plotone{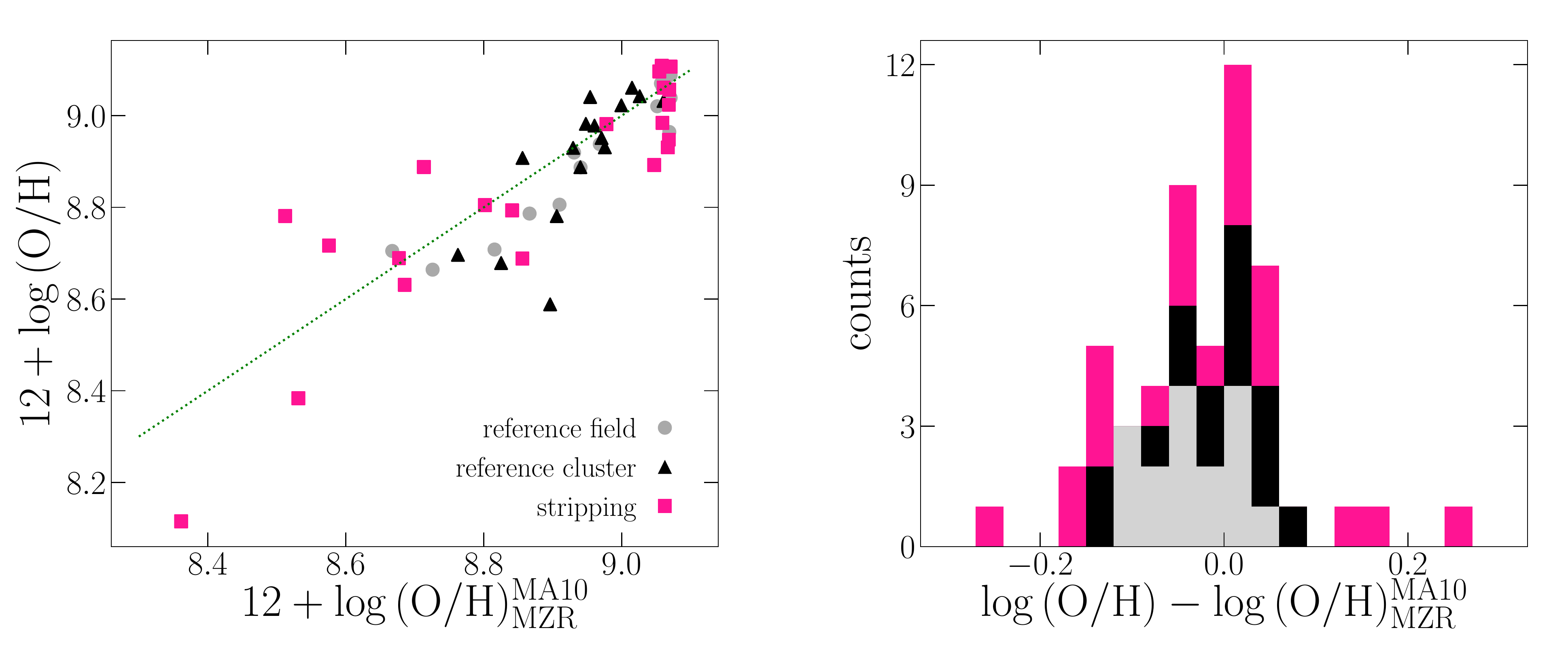}
\caption{{\em Upper panel}: Distribution of the galaxy metallicities within a diameter of 5~kpc and the expected value according to the Eq.~2 in \citet{mannucci2010} ({\em on the left}) and the stacked histogram of the differences ({\em on the right}). {\em Lower panel}: As the upper panel, but considering Eq.~1 in \citet{mannucci2010}. The gray circles, black triangles and pink squares indicate the reference field, reference cluster and stripping galaxies, respectively. The histograms follow the same color-code.\label{fig:deltaFMR}}
\end{figure*}

In Fig.~\ref{fig:FMRfiber} we show the stellar mass-SFR-metallicity space with our galaxies superimposed on the FMR surface in \citet[Eq.~2]{mannucci2010}. Our sample is well distributed along the analytic relation although it is worth to make some considerations. Despite the existence of a narrow total SFR-mass relation in the GASP sample \citep{vulcani2018b}, the  SFR values measured inside the aperture manifest a large dispersion with the stellar mass, in particular we observe some galaxies with low SFR at high masses. On the other hand, our sample does not contain galaxies with high SFR at low masses, so we are lacking data in the region where the FMR is the most sensitive to the SFR.

To quantify the scatter around the surface we compare the gas-phase metallicity of our datapoints and the expected value from the FMR \citep[Eq.~2]{mannucci2010} given the mass and the SFR.  In the top panels of Fig.~\ref{fig:deltaFMR} we show this comparison and the distribution of the differences. The gas-phase metallicity inferred inside the aperture are on average lower by $0.041$~dex than the expected values, and the scatter amounts to 0.091~dex. The distribution is slightly asymmetric with a skewness $s=-0.21$. The shift of the metallicities towards lower values than those expected from the FMR of \citet{mannucci2010} could be due to the fact that we adopted the metallicity calibrator based on the R3 index instead the R23 one, as \citeauthor{mannucci2010} did.

In the bottom panels of Fig.~\ref{fig:deltaFMR} we compare instead our values with the expected values from the MZR curve of \citet[Eq.~1]{mannucci2010} given the stellar mass. We again note that most of galaxies are mainly located below the considered relation. The distribution of the residuals is centered at $-0.029$~dex with a dispersion of 0.094~dex, and results more symmetric ($s=-0.007$) than the residual distribution derived considering the FMR. At high masses we observe a saturation of the expected values due to the flattening of the function at the high-mass end.
We infer that our sample certainly follows both relations proposed by \citet{mannucci2010}, however the scatter of our galaxies along the MZR is reduced of only 0.003 when the SFR is taken into account. We stress again that our sample does not contain galaxies with low stellar mass and high SFR. Moreover, we note that in the plane ($\mathrm{\log\,M}$,$\mathrm{\log\,SFR}$) sampled by our galaxies the MZR and the FMR of \citet{mannucci2010} are similar, and in fact the metallicities of our galaxies derived by the two relations differ on average of only 0.12 dex, with a scatter of 0.04 dex.

We obtain similar results also comparing our O3N2-based $\langle {\rm O/H} \rangle_{<0.5\,R_{\rm e}}$ values, total SFRs, and galaxy stellar masses with the analytical form of the FMR presented by \citet[Eq.~5]{curti2020}. The distribution of the differences along the metallicity-axis between our datapoints and this surface almost peaks at zero (with a mean of -0.03~dex and a dispersion of 0.08~dex) although it is highly skewed ($s=-1.52$).

In conclusion, both the analysis of residuals of the gas metallicity at the $R_{\rm e}$ along the MZR using the spatially resolved data (Fig.~\ref{fig:MZR_SFR}) and the comparison of the values inside the aperture of 5~kpc with the FMR of \citet{mannucci2010} (Figs.~\ref{fig:FMRfiber} and \ref{fig:deltaFMR}) are in agreement with the possibility of a secondary dependence of the MZR on the SFR. However, this dependence is relevant only when including the very low mass galaxies.
A limitation of our analysis comes from the low number statistics and from the fact that the galaxies, here studied, follow a tight mass-SFR relation \citep{vulcani2018b}: at a given mass the galaxies span too small a SFR range to appreciate in detail the dependence on the SFR. Above $10^{9.25}~{\rm M_\odot}$ the gas metallicities of our galaxies is mainly driven by the stellar mass and it is not necessary to introduce the SFR as third parameter to explain their distribution. On other hand, the presence of low mass galaxies with higher metallicity than the common MZR testifies a more complex picture.

\subsection{Dependence on environment}\label{sec:env}

\begin{figure}
\centering
\includegraphics[width=\columnwidth]{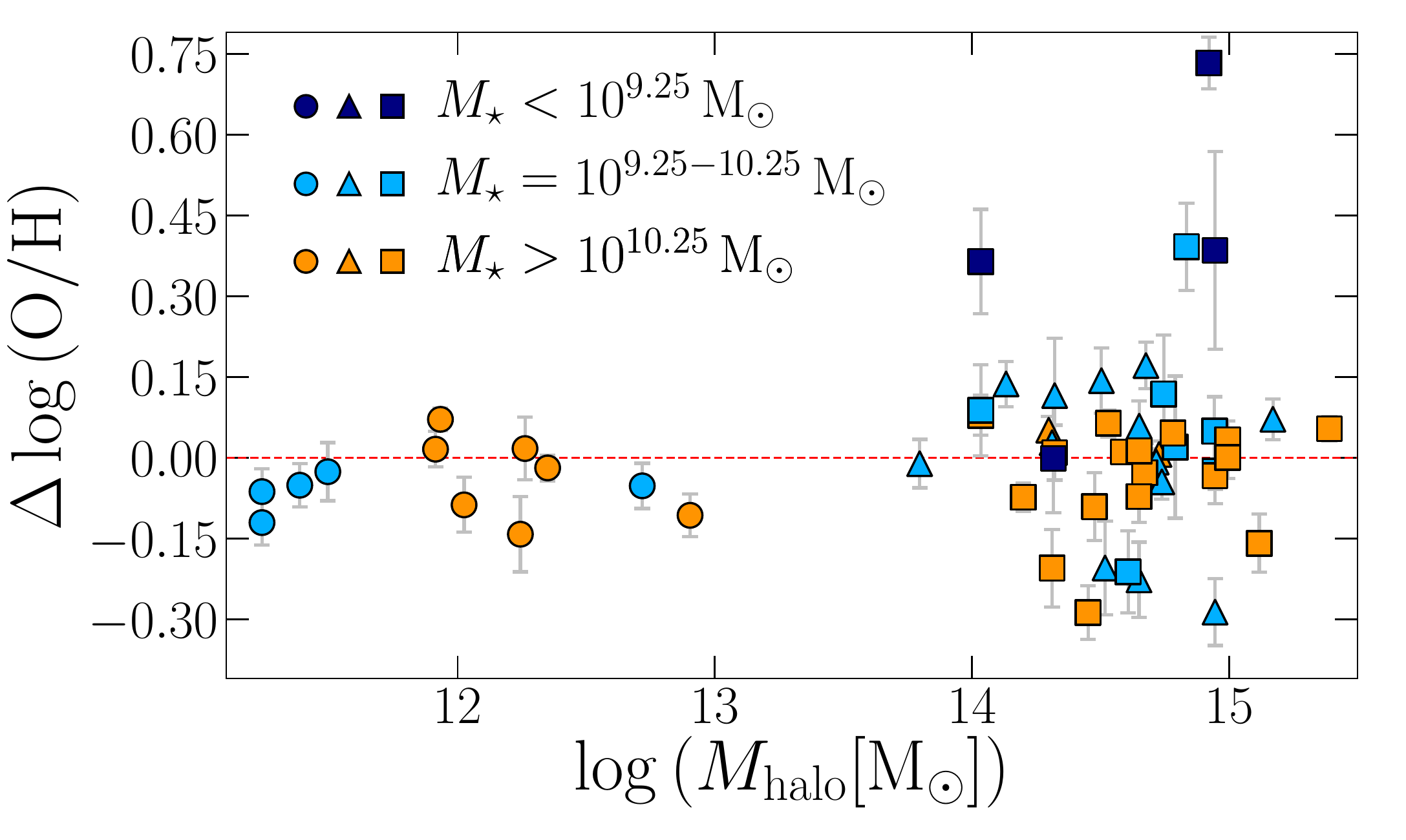}
\includegraphics[width=\columnwidth]{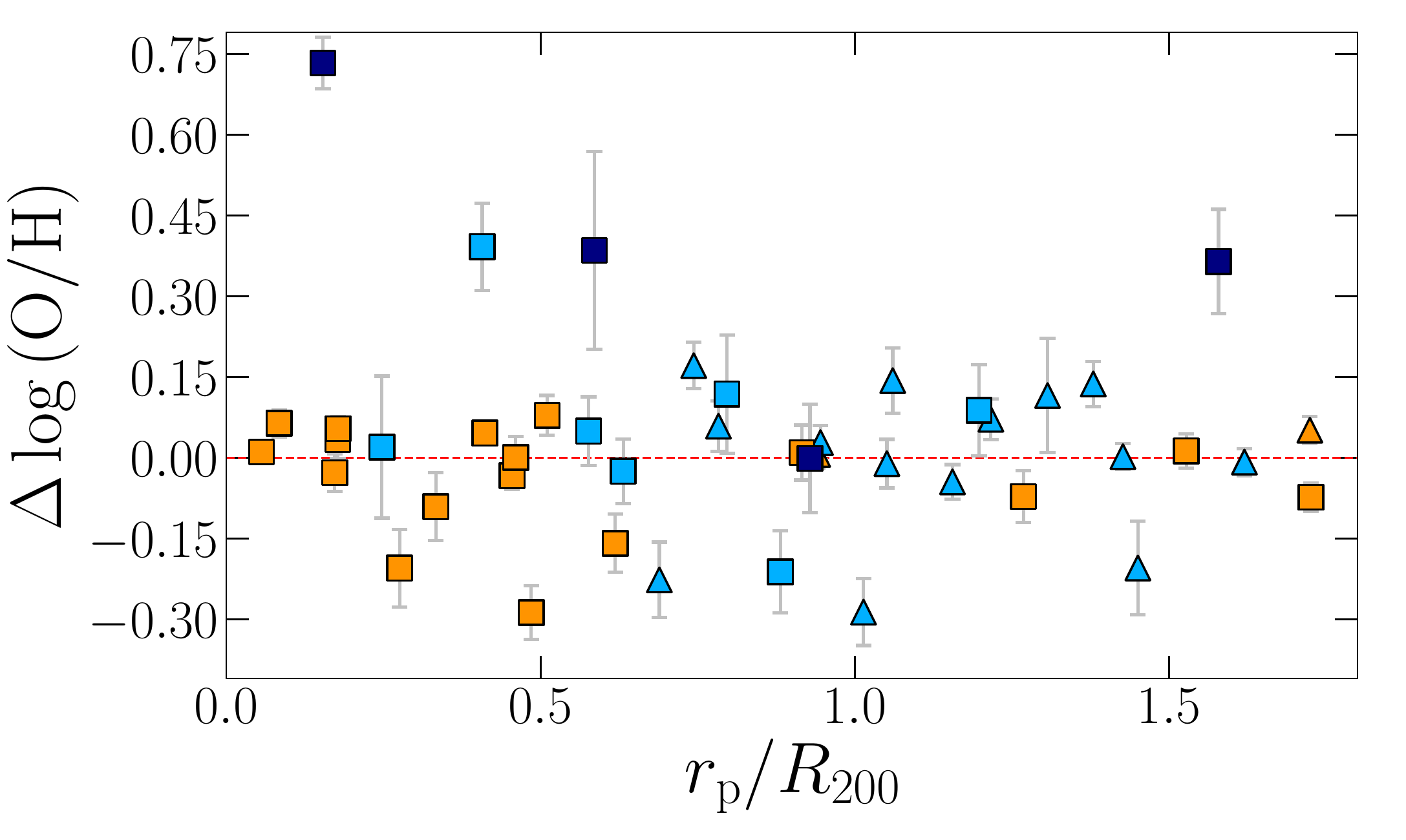}
\includegraphics[width=\columnwidth]{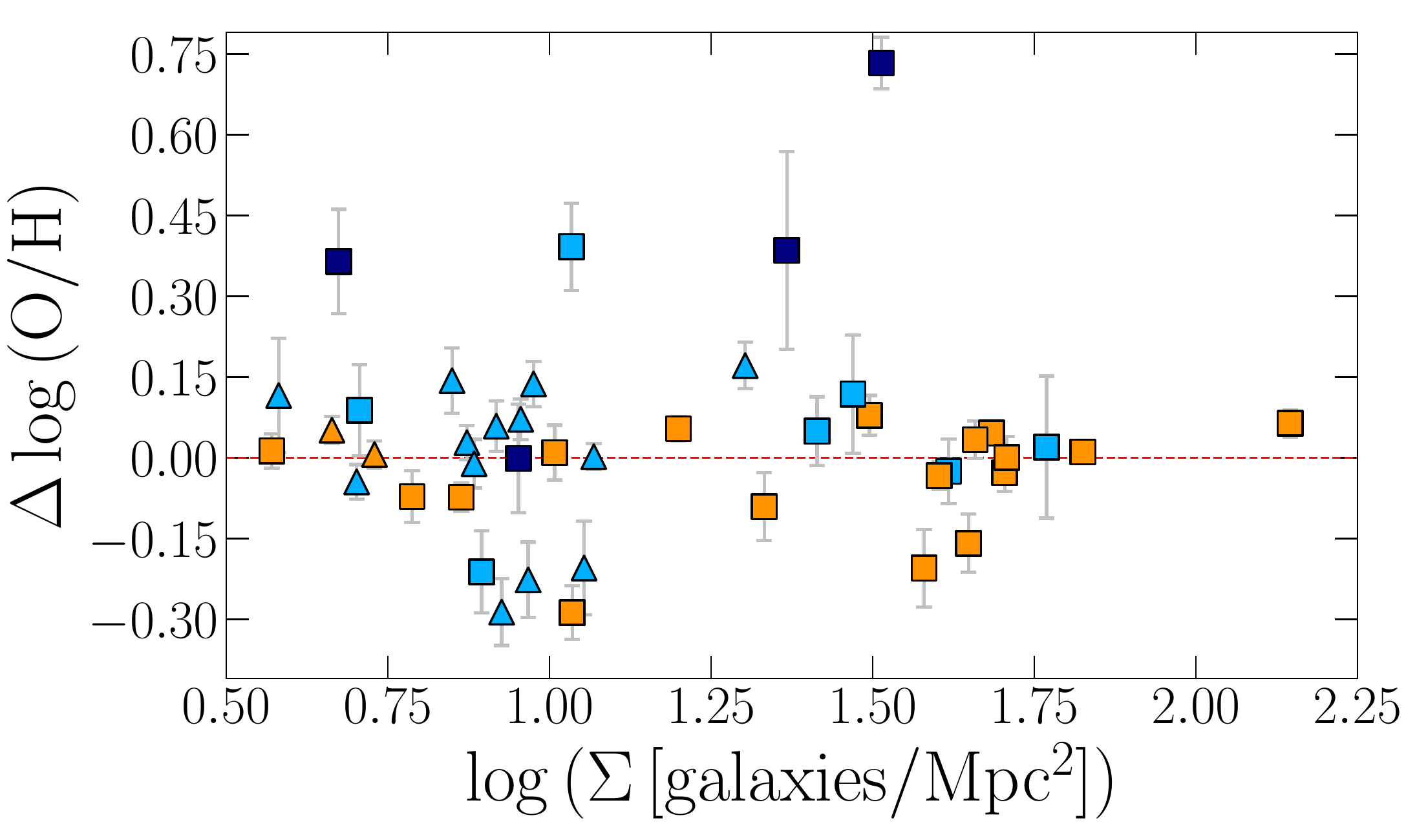}
\caption{Residuals of the MZR versus $M_{\rm halo}$ ({\em upper panel}), versus the projected clustercentric distance ({\em middle panel}) and versus the projected galaxy density ({\em lower panel}). The middle and bottom panels show only the cluster galaxies. Circles, triangles and squares indicate the reference field, reference cluster and stripping galaxies, respectively. Colors refer to the galaxy stellar masses as in the legend. The horizontal red line corresponds to the MZR of Fig.~\ref{fig:mzrpyqz}.}
\label{fig:MZR_env}
\end{figure}

In Fig.~\ref{fig:binnone} we have already observed that
intermediate mass field galaxies have on average lower gas metallicity than cluster galaxies of same mass. Instead, at high masses they follow a similar MZR. Reference cluster galaxies and stripping galaxies do not show relevant offset in metallicity but are more spread along the MZR than the field galaxies. Now we attempt to improve the analysis and investigate the relation between the gas-phase metallicity residuals and the properties of the host environment.

As previously discussed, the galaxies with the largest metallicity residuals (JO149, JW56, JO45 and JO162) are cluster galaxies with $M_\star<10^{9.5}\,{\rm M}_\odot$, in agreement with the results of \citet{pilyugin2017} who find overabundances for low mass galaxies ($M_\star<10^{9.6}\,{\rm M}_\odot$) in the most crowded environments.

The upper panel of Fig.~\ref{fig:MZR_env} shows the distribution of the MZR residuals as a function of the mass of the host halo ($M_{\rm halo}$). The halo masses of the groups are taken from \citet{paccagnella2019}, while those of the clusters from \citet{biviano2017} and Munari et al.\ (in prep.), and listed in Gullieuszik et al.\ (submitted). No significant trends are detected, but we can appreciate a larger scatter for the cluster galaxy sample (0.18~dex), mainly due to low mass galaxies, than for the sample of the field galaxies (0.08~dex). 

In the middle panel of Fig.~\ref{fig:MZR_env} we compare the metallicity residuals of the cluster galaxies with the projected clustercentric distance $r_{\rm p}/R_{200}$. Also in this case we do not observe any correlation.

For cluster galaxies, we also take into account the galaxy local density. Several studies, based on thousands of galaxies, found hints of a correlation between the gas metallicity and the local density, meaning that the local density has only a marginal role on the chemical evolution of the galaxies \citep{ellison2009,peng2014,pilyugin2017}.

We adopt here the projected density $\Sigma=N/A$\footnote{$A$ is the circular area that encloses the $N$-th nearest galaxy neighbour. Here, we adopted $N=10$.} already exploited by \citet{vulcani2012} to characterize the galaxies of the WINGS cluster survey, and calculate the values for the OMEGAWINGS galaxies (Vulcani 2019, priv. communication).
The lower panel of Fig.~\ref{fig:MZR_env} depicts the distribution of the residuals as a function of the local projected galaxy density. No significant trends are observed.

Finally, even when we take into account two quantities together, such as the halo mass at fixed clustercentric distance (plot not shown), no significant trend arises.

It is however possible that our analysis
based on a relatively small sample has not a sufficient sensitivity to appreciate the environmental effects on the metal enrichment.

Nonetheless, we do observe a large scatter of the residuals for the cluster galaxies, especially the low mass ones, suggesting that these objects could experience several processes in a dense environment, able to more efficiently alter the metallicity of the gas than those in less dense environments.
In particular, we report the case of the low mass JW56 stripping galaxy, whose metallicity residual than the MZR is the largest one in the sample. This galaxy is within a high massive galaxy cluster ($M_{\rm halo}=10^{14.9}\,{\rm M_\odot}$), close to the center ($r_{\rm p}/R_{200}=0.16$) and at a high local galaxy density ($\Sigma=10^{1.51}\,{\rm galaxies\,per\,Mpc^{2}}$). Therefore JW56 lives in an extreme condition that not only facilitates the RPS but also could prevent the accretion of metal-poor gas and be responsible for its high metallicity.

\subsection{Dependence on the galaxy size}\label{sec:paramdep}

\begin{figure}
\centering
\includegraphics[width=\columnwidth]{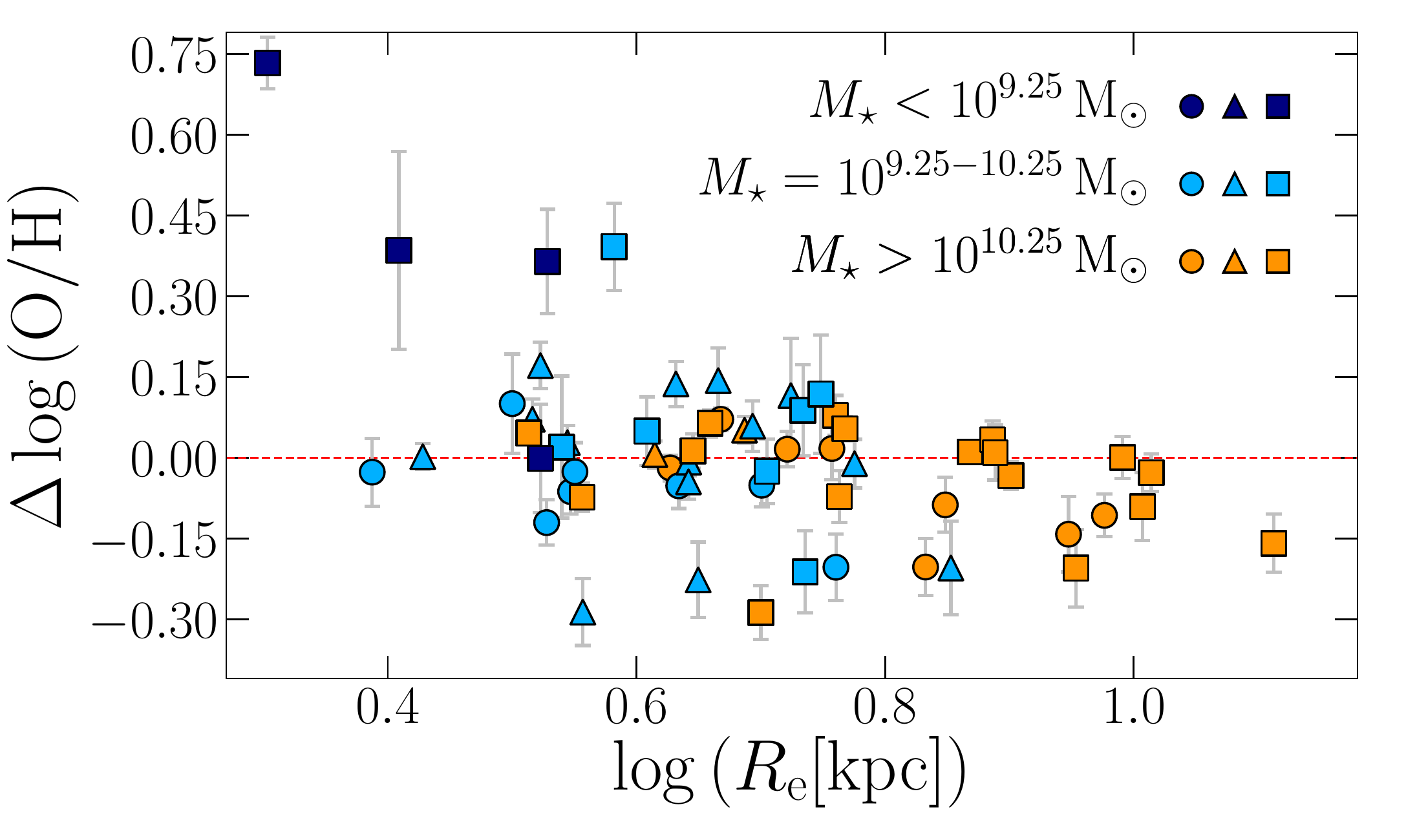}
\caption{Residuals of the MZR versus $R_{\rm e}$. Circles, triangles and squares indicate the reference field, reference cluster and stripping galaxies, respectively. Colors refer to the galaxy stellar masses as in the legend. The horizontal red line corresponds to the MZR of Fig.~\ref{fig:mzrpyqz}.}
\label{fig:MZR_par}
\end{figure}

Observational studies and cosmological numerical simulations observed that at a given mass smaller galaxies are on average metal richer \citep{ellison2008,sanchezalmeida2018}.

In Fig.~\ref{fig:MZR_par} we investigate the possible correlation with the size of the galaxies. For galaxies with $M_\star>10^{9.25}$~M$_\odot$ the metallicity residuals only show a mild anti-correlation with the $R_{\rm e}$ ($r[d.f.=54]=-0.29$, $p=0.03$). If we include also the residuals of the lower mass galaxies, the trend becomes more evident ($r[d.f.=58]=-0.47$, $p=0.0002$). In particular, we highlight that the galaxy with the largest metallicity residual (JW56) has the smallest effective radius of the sample, resulting more compact than galaxies with similar mass.

We stress that the $R_{\rm e}$ of JO149 and JO95 are estimated from the mass-size relation described in Appendix~\ref{app:re}, but even if we exclude these two objects the anti-correlation remains pronounced with the same level of significance.
The $R_{\rm e}$ of the stripping galaxies could be overestimated because of the light enhancement in the external regions due to the recent star formation. Nonetheless, we do not detect significant offsets between the effective radii of the various categories (see Appendix~\ref{app:re} for detail).

\section{Summary and Conclusions}\label{sec:concl}
In this paper, for the first time, we studied the gas-phase metallicity relation in cluster galaxies undergoing RPS observed within the ESO MUSE GASP Program. These galaxies were selected for showing evident signs of only-gas removal due to the interaction between the ICM and their ISM. 

We explored the MZR of these peculiar objects in the stellar mass range $10^{8.8}\le M_\star \le 10^{11.5}\,{\rm M_\odot}$ and compared it with a reference sample of galaxies, both in clusters and in the field, with no significant signs of ongoing gas stripping processes.

To characterize the global metallicity of these galaxies, we used the {\sc pyqz} code and derived the mean gas-phase metallicity of the ionized gas at the $R_{\rm e}$. Our relative abundances are consistent with those inferred using the empirical calibration of \citet{curti2017} based on the O3N2 index. 

Both the stripping and the reference samples follow the same well-defined MZR down to $10^{9.25}\,{\rm M_\odot}$ with a scatter of 0.12~dex.
At $M_\star<10^{10.25}\,{\rm M_\odot}$, the field galaxies show on average lower metallicities than the cluster galaxies, both stripping and reference, with a significance level $>$1$\sigma$, while at high masses the offset disappears. No differences are detected between reference cluster galaxies and those undergoing RPS, but large scatters are observed, mainly at low masses.
We detect
4 stripping galaxies with low masses ($<10^{9.4}\,{\rm M_\odot}$), that have high gas metallicity with the largest differences ($>$0.36~dex) from the observed MZR. Their overabundance does not appear to be connected to the redistribution of the gas in the disk due to the RPS and the importance of a fast self-enrichment as a consequence of the recent SFR enhancement by the RPS is not clear.
 
The scatter around the MZR can be explained by the interconnection with the physical properties of these galaxies. Indeed,  results are consistent with a secondary dependence on the SFR and on the galaxy size, even if the anti-correlation between gas-phase metallicity and these parameters is mainly driven by very low mass galaxies. In particular, JW56, the stripping galaxy with $M_\star=10^{9.05}\,{\rm M_\odot}$ and the largest overabundance in the sample (0.73~dex), is more compact than galaxies with similar mass and has a very low SFR.

We did not find any correlation between the gas-phase metallicity and some environmental properties (halo mass, projected clustercentric distance, local galaxy density). However the location of JW56, close to the center of a massive cluster, could play a role in the metal enrichment of this galaxy preventing the accretion of metal-poor gas, in addition to foster the RPS.

Our analysis based on the mean gas-phase metallicity at the ${R_{\rm e}}$ did not highlight a dependence on the RPS process.
Therefore, either the RPS does not alter the metal content around ${R_{\rm e}}$ (at least until the inner regions of the disk get stripped), or taking the mean of the metallicity values around the ${R_{\rm e}}$ could prevent to appreciate the displacement of the 
gas inside the disk with respect to the galaxy center and the consequent lopsidedness of the radial gas metallicity distribution.

To investigate how the RPS works, a detailed study of the spatially-resolved gas-phase metallicity (inside the disk and along the ionized gas tails) is currently ongoing and  will be presented in a forthcoming paper.

\acknowledgements
We thank Stephanie Tonnesen for useful discussions. We thank the anonymous referee for their comments that have improved the paper. Based on observations collected at the European Organization for Astronomical Research in the Southern Hemisphere under ESO programme 196.B-0578. This project has received funding from the European Research Council (ERC) under the European Union's Horizon 2020 research and innovation programme (grant agreement No.~833824). We acknowledge funding from the INAF main-stream funding programme (PI B.~Vulcani). B.~V., M.~G.\ and D.~B.\ also acknowledge the Italian PRIN-Miur 2017 n.20173ML3WW\_001 (PI A.~Cimatti).
J.~F.\ acknowledges financial support from the UNAM- DGAPA-PAPIIT IA103520 grant, M\'exico. Y.~J.\ acknowledges financial support from CONICYT PAI (Concurso Nacional de Inserci\'on en la Academia 2017) No.~79170132 and FONDECYT Iniciaci\'on 2018 No. 11180558. We acknowledge financial contribution from the contract ASI-INAF n.2017-14-H.0.

\facility{VLT(MUSE)}
\software{{\sc sinopsis}, {\sc kubeviz}, {\sc iraf}, {\sc pyqz}, {Python}}

\appendix
\section{Surface brightness analysis}\label{app:re}

Studying the oxygen abundance at different distances from the galactic center requires to know the structural parameters of each galaxy.

The effective radius ($R_{\rm e}$) of the galaxies is derived by the analysis of the azimuthally-averaged surface brightness profile (SBP), while the position angle ($P\!A$), ellipticity ($\varepsilon$) and the inclination ($i$) are estimated from the disk isophotes.

The procedure we adopted exploits the I-band images produced by the MUSE pipeline integrating the reduced datacubes with the I-band filter response curve. Among the available filters entirely included in the wavelength range of MUSE, the I-band samples the reddest part of the spectrum. It therefore  yields a smoother luminosity distribution, 
less affected by the youngest stars, formed during recent star formation episodes, and by peaks of ionized gas emission produced by these stars. In fact, these two light sources are
scattered along the galaxy disk and arise over the light of the less massive and older stars, that are more homogeneously distributed. At the redshift of the GASP galaxies, the I-band avoids to pick up the strong emission of the H$\alpha$ and [{\sc O\,iii}], even if the lines of the [{\sc S\,ii}] always fall in this spectral range.
For the same reason the stripped gas tails appear fainter in the I-band images, thus preventing
the SBP to be
biased by the stars formed in the tails due to ram pressure stripping. Furthermore we can better observe the galactic bars that are more frequently seen at red and near-infrared wavelengths \citep{knapen2000}. The downside is that the sky subtraction on the red part of the MUSE spectra is less effective and we detect more sky noise than in the blue part.

We perform an isophotal analysis on the I-band images using the {\sc ellipse} task in {\sc iraf} \citep{jedrze1987} to extract the SBP of the galaxies. {\sc ellipse} fits on the image a series of elliptical isophotes such to minimize the deviations from the real shape of the galaxy isophotes. Then it returns the mean intensity along the ellipse, semi-major axis, $P\!A$ and $\varepsilon$ for each one. We mask out foreground stars, nearby and background galaxies and bad pixels before fitting the isophotes. We also mask the bright clumps on the galaxy disk that do not follow the SBP of the disk. They mainly correspond to star forming regions and spiral arms.

\begin{figure}
\centering
\includegraphics[width=0.45\textwidth]{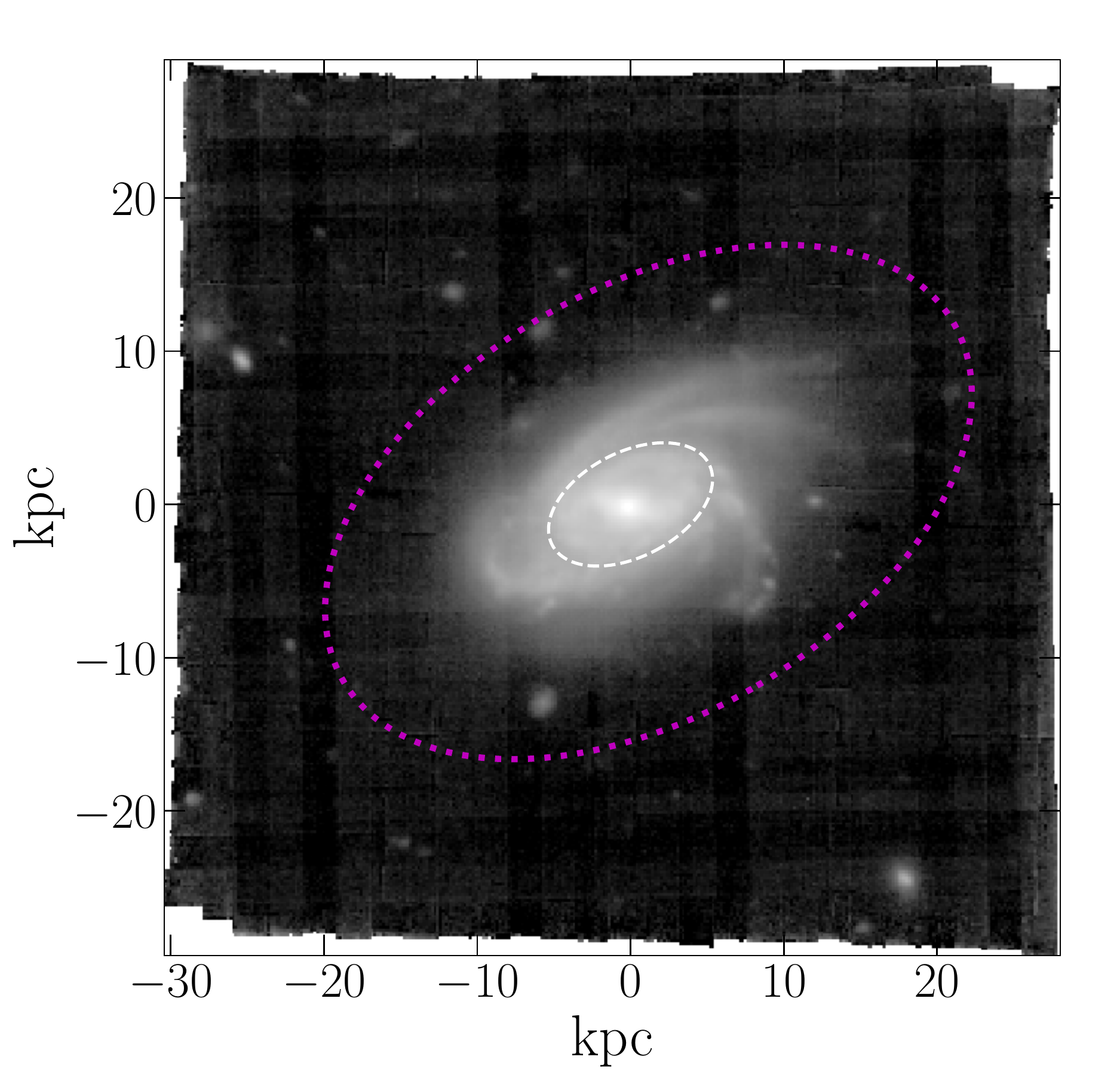}
\includegraphics[width=0.45\textwidth]{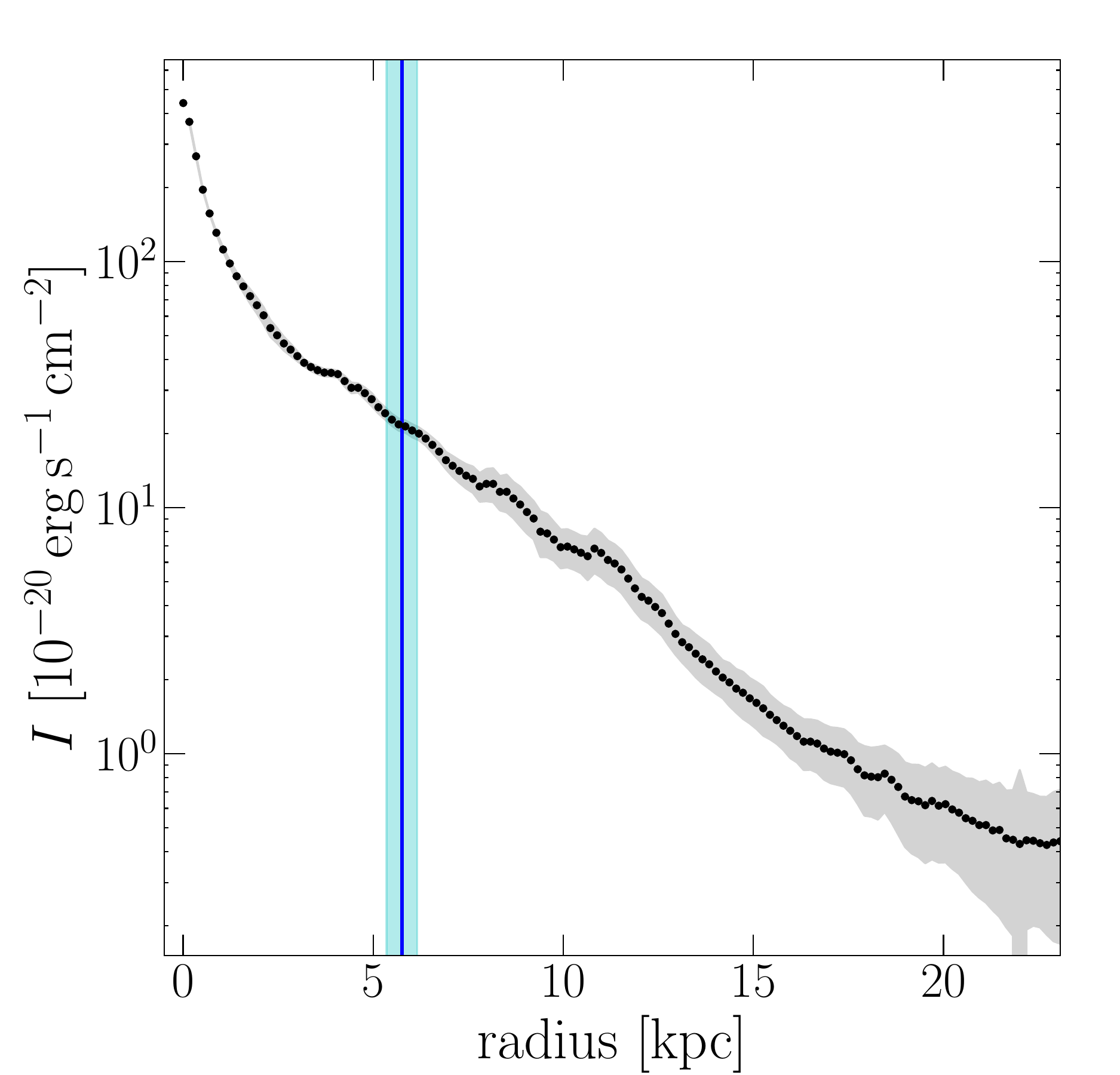}
\caption{{\em Left panel:} I-band image of JO49 galaxy. The white inner ellipse indicates the $R_{\rm e}$ of the galaxies according to the mean $\varepsilon$ and the mean $P\!A$ of the disk. The magenta ellipse corresponds to the outermost elliptical isophote fitted by the {\sc ellipse} algorithm. {\em Right panel:} Radial profile of the surface brightness along with the rms errors of JO49 extracted by {\sc ellipse}. The vertical blue line and the cyan area indicate the inferred $R_{\rm e}$ and the corresponding error.}\label{fig:SBP}
\end{figure}

Taking advantage of the wide field-of-view of the GASP data, we measure the radial profile of the surface brightness beyond the detectable extent of the galaxy to probe the surface brightness of the background. After having checked that the extracted SBP flattens out at large galactocentric distances we compute the mean intensity of the unmasked pixels outside the elliptical isophotes. The inferred value is comparable with the intensity of the last isophotes and represents a residual sky level of the image. Therefore we subtract this value from the SBP and proceed to estimate the $R_{\rm e}$.

We derive the luminosity growth curve as the trapezoidal integral
\begin{equation}
L(R) = 2\pi \int^{R}_{0} I(r)\bigl(1-\varepsilon(r)\bigr)\,r\,dr,
\end{equation}
where $I(r)$ is the SBP, $\varepsilon(r)$ is the $\varepsilon$ profile and $r$ is the semi-major axis of the elliptical isophotes. By definition the $R_{\rm e}$ is the radius that encloses half of the total luminosity $L_{\rm tot}$. We approximate $L_{\rm tot}\approx L(r_{\rm max})$, where $r_{\rm max}$ corresponds to the last fitted isophote containing the whole extent of the galaxy observed in the MUSE data and compute $R_{\rm e}$ such that $L_(R_{\rm e})/L_{\rm tot}=0.5$. In addition, in order to derive the upper and lower limits of the $R_{\rm e}$ we repeat the computation using $(I\pm\sigma)(r)$ within the integral, where $\sigma$ is the error of the mean isophote intensity.

Lastly, using the isophotes that trace the galaxy disk we calculate their mean $P\!A$ and $\varepsilon$.

As an example, Figure~\ref{fig:SBP} shows the I-band image of the JO49 galaxy and the SBP extracted using {\sc ellipse}. The magenta ellipse, traced on the image, corresponds to the outermost elliptical isophote fitted by the algorithm. We observe that the detectable extension of the galaxy is enclosed inside this isophote, also confirmed by the flattening of the SBP at the largest radius.

\subsection{Mass-size relation}

\begin{figure*}
\plotone{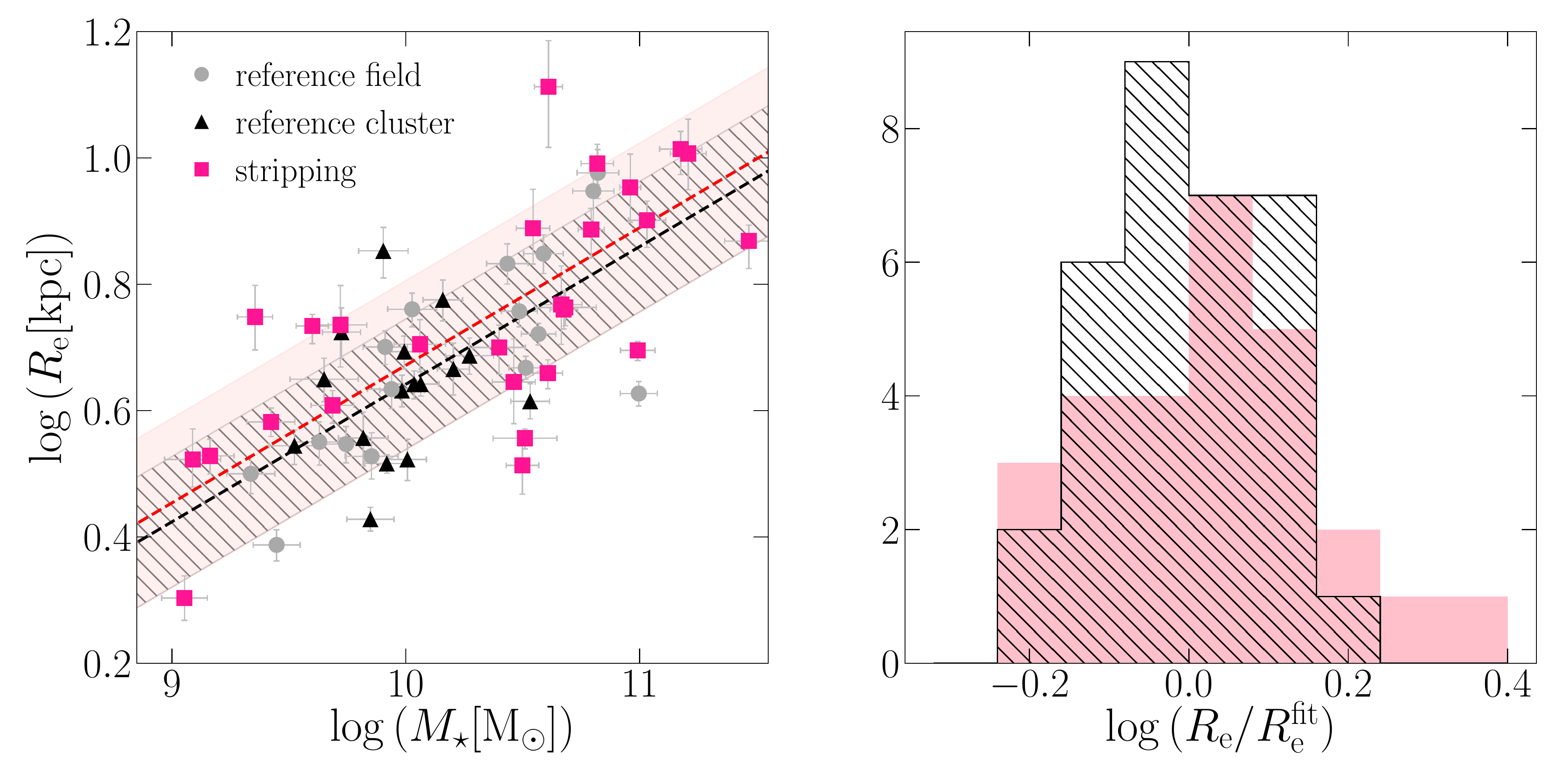}
\caption{{\em Left panel}: Relation between the galaxy stellar mass and the $R_{\rm e}$. The gray circles, black triangles and pink squares indicate the  reference  field, reference cluster and stripping galaxies, respectively. The black dotted line and the dashed gray area show the best linear fit and the rms error for the reference sample. The red dotted line and the pink area show the best linear fit and the rms error for the stripping galaxy sample, adopting the same slope as the reference sample fit. {\em Right panel}: Distribution of the differences between the estimated effective radii and the expected values according to the best fit of the reference sample, given the stellar mass. The black dashed area corresponds to the reference sample, the pink area refers to the stripping galaxy sample. \label{fig:mre}}
\end{figure*}

We investigate the mass-size relation of the sample, comparing the stellar mass with  $R_{\rm e}$, shown in the left panel of Fig.~\ref{fig:mre}.
We observed a well-established correlation between the quantities, both for the reference and stripping galaxies. We explore the hypothesis that the stripping can alter the measurement of the $R_{\rm e}$ due to a possible enhancement of the luminosity in the external regions of the galaxies. To examine the differences between the two samples, we fit the data with a linear regression based on a least square fitting method. For the reference sample we allow the slope and the intercept to vary, while for the stripping sample we assume the same slope of the reference sample. The mass-size relations of the reference and the stripping galaxies are described by the following equations
\begin{align}
\log\,R_{\rm e} = (0.218\pm 0.002)\,\log\,M_\star+(-1.536\pm0.233) & \qquad\text{(reference galaxies)}\label{eq:mrecontr}\\
\log\,R_{\rm e} = 0.218\,\log\,M_\star+(-1.506\pm0.001) & \qquad\text{(stripping galaxies)}\label{eq:mrestrip}
\end{align}
with a scatter of 0.10 and 0.13~dex, respectively. The difference between the two fits is $\sim$0.030~dex. 
Differences are better seen in the right panel of Fig.~\ref{fig:mre}, that shows the distribution of the difference between the $R_{\rm e}$ of each galaxy and the value derived from the reference sample fit given the galaxy mass. Although there's a small tail towards higher values of the $R_{\rm e}$ for the stripping galaxies, the KS test cannot reject the hypothesis that the two distributions are drawn from the same parent distribution (p-value $\sim$0.28).

\bibliography{myref}

\end{document}